\documentclass[journal]{IEEEtran}
\usepackage{graphicx}
\usepackage{amsmath}
\usepackage{amssymb}
\usepackage[noadjust]{cite}

\usepackage{array} 
\usepackage{algorithm,algpseudocode}
\newtheorem{theorem}{Theorem}[section]

\newtheorem{lemma}[theorem]{Lemma}
\newtheorem{remark}{Remark}

\ifCLASSINFOpdf
\else
\fi
\begin{document}
%
% paper title
% Titles are generally capitalized except for words such as a, an, and, as,
% at, but, by, for, in, nor, of, on, or, the, to and up, which are usually
% not capitalized unless they are the first or last word of the title.
% Linebreaks \\ can be used within to get better formatting as desired.
% Do not put math or special symbols in the title.
\title{Time-Varying Formation Control of a Collaborative Multi-Agent System Using Negative-Imaginary Systems Theory}
%
%
% author names and IEEE memberships
% note positions of commas and nonbreaking spaces ( ~ ) LaTeX will not break
% a structure at a ~ so this keeps an author's name from being broken across
% two lines.
% use \thanks{} to gain access to the first footnote area
% a separate \thanks must be used for each paragraph as LaTeX2e's \thanks
% was not built to handle multiple paragraphs
%

\author{Vu Phi Tran,~Matthew~Garratt and~Ian~R.Petersen%
	\thanks{Vu~Phi~Tran and~Matthew~Garratt are with the School of Engineering and Information Technology, University of New South Wales, Australia and~Ian~R.Petersen is with the Research School of Engineering, Australian National University, Australia. {Phi.Tran}@student.adfa.edu.au, {M.Garratt}@adfa.edu.au and {I.R.Petersen}@gmail.com}
}

\maketitle
% As a general rule, do not put math, special symbols or citations
% in the abstract or keywords.
\begin{abstract}
The movement of cooperative robots in a densely cluttered environment may not be possible if the formation type is invariant. Hence, we investigate a new method for time-varying formation control for a group of heterogeneous autonomous vehicles, which may include Unmanned Ground Vehicles (UGV) and Unmanned Aerial Vehicles (UAV). We have extended a Negative-Imaginary (NI) consensus control approach to switch the formation shape of the robots whilst only using the relative distance between agents and between agents and obstacles. All agents can automatically create a new safe formation to overcome obstacles based on a novel geometric method, then restore the prototype formation once the obstacles are cleared. Furthermore, we improve the position consensus at sharp corners by achieving yaw consensus between robots. Simulation and experimental results are then analyzed to validate the feasibility of our proposed approach.
\end{abstract}

% Note that keywords are not normally used for peerreview papers.

% For peer review papers, you can put extra information on the cover
% page as needed:
% \ifCLASSOPTIONpeerreview
% \begin{center} \bfseries EDICS Category: 3-BBND \end{center}
% \fi
%
% For peerreview papers, this IEEEtran command inserts a page break and
% creates the second title. It will be ignored for other modes.
\IEEEpeerreviewmaketitle

\section{Introduction}
% The very first letter is a 2 line initial drop letter followed
% by the rest of the first word in caps.
% 
% form to use if the first word consists of a single letter:
% \IEEEPARstart{A}{demo} file is ....
% 
% form to use if you need the single drop letter followed by
% normal text (unknown if ever used by the IEEE):
% \IEEEPARstart{A}{}demo file is ....
% 
% Some journals put the first two words in caps:
% \IEEEPARstart{T}{his demo} file is ....
% 
% Here we have the typical use of a "T" for an initial drop letter
% and "HIS" in caps to complete the first word.
\IEEEPARstart{M}{ulti-vehicle} coordination has been attracting increasing attention from researchers as it provides much enhanced capability for applications such as search and rescue or mapping. Cooperation between multiple vehicles can lead to faster and more effective missions. For example, a cooperating UAV can fly above obstacles and travel faster than a UGV, whilst providing sensor coverage for a much wider area than that accessible to the UGV. Meanwhile the UGV can inspect the environment more closely and precisely than the UAV, which flies at altitude and may have a partially blocked view. Supporting ground transportation task is a potential application of the UAV-UGV cooperation \cite{Forster,VIdal}. In this task, a UAV can be fitted with a camera to capture the visual pose of UGVs moving on the ground, helping to localize the UGVs and guide them along a safe path. In summary, the cooperation of UAVs-UGVs is beneficial in many autonomous missions.

There are two control strategies for a multi-vehicle system: centralized or decentralized. In the former approach, a ground station (GS) collects all required data (posture, velocity, acceleration) and transmits proper control signals to each agent. However, centralized control depends on perfect communication which can be an issue if data dropouts occur due to overloaded networks or when communication range constraints are challenged \cite{Milutinovic06,Huang12}. With decentralized control, a local controller is designed for each robot, and the control signals are provided by only using local information about agents and their neighbors \cite{Sahraei10,deLope13}. This approach significantly reduces the amount of transmission data, time delay and is more robust against communication failures. Therefore, we base our consensus control approach on a decentralized strategy where the agents know only their own velocities, accelerations, Euler angles, and their relative position from other agents.

Within the application of multi-vehicle systems, forming and maintaining a prescribed formation is a real challenge. A large body of work related to multi-robot systems can be found in the literature \cite{Yan,Dong1, Dong2,Su}. More consensus results are shown in \cite{Dong1,Dong2,Qu,Seo}. These papers pay much attention to formation stabilization problems, but the internal stability of each vehicle based on these cooperative methodologies is not sufficiently addressed. Furthermore, as seen in \cite{Dong3,Wang1,Wang2,Dong2}, agents in a cooperative network are often treated as second-order systems to solve the formation tracking problems. However, the position and velocity tracking loops of the UAV or the UGV are actually represented by a higher order transfer function. For example, the closed-loop velocity control of a quadrotor leads to at least a fourth-order model \cite{Li11,Zhang14}. Using an improvement in NI consensus theory, our method can solve this problem by considering simultaneously the stability of the formation control and dynamics of agents.

More and more researchers have realized that formation control problems can be handled using consensus-based approaches. For example, the relevant works in \cite{Turpin12, Tran17} proposed time invariant formation controllers for multiple UAVs. However, in most practical tasks, multi-robot systems should also be able to vary their formation over time so that they can make progress towards the goal position while avoiding obstacles in an unknown environment.

The shortcoming of using a fixed formation has been overcome recently by a few researchers. In \cite{Dong1, Dong2}, a time-varying formation controller was tested and validated on five quadrotors. For this method to work, each agent must be able to sense the absolute position and orientation of its neighbor. As a result, time delays or data drop during inter-agent communication may occur as the number of agents increases dramatically. Similarly, a time-varying formation control approach for a UAV based on a virtual leader structure is illustrated in \cite{Rahimi14}, where the follower vehicles maintain a relative position from a simulated leader represented by an idealized trajectory. Compared to these methodologies \cite{Dong1, Dong2, Rahimi14}, our slave agents only utilize measurements of the relative distances from their master and each other, such that no data need to be exchanged during the concensus process.    

One of the practical applications of time-varying formation control approaches for avoiding unexpected obstacles, has been considered in \cite{Rahimi14, Lee18, Dai15}. Once a potential collision between the robots and an obstacle is predicted, the robots in \cite{Rahimi14} will automatically produce a queuing formation for safely avoiding the corresponding obstacle. However, this strategy cannot be used to avoid multiple obstructions. This issue was solved in \cite{Lee18}. Whenever an obstacle is detected inside the repulsion zone, an escape angle is determined by the robot orientation relative to the real target and the disposition of ultrasonic sensors in the ring, in relation to the axis of robot movement. It is pointed out from the simulation results that the stability of the obstacle avoidance algorithm was not examined. Besides, the control output command pairs based on the behavioral control approach, including turn left/turn right and forward/backward, coincide in some cases since the working area of each control layer is not adaptive. A better formation control strategy for obstacle avoidance was developed by Dai, Y., et al \cite{Dai15} in 2015. In this work, a safe path for the leader was planned by a Geometric Obstacle Avoidance Control Method (GOACM) where the slaves are moved into a safe formation relative to the leader. Unfortunately, because of sudden changes in the velocity setpoints and control saturation while switching into a safe triangle formation, this method suffers from oscillations in robot position. Moreover, this method is based on the real surface boundary of an obstacle to formulate the rotation angle, making it not applicable for practical situations where obstacles often have arbitrarily complicated shapes.

\begin{figure}
	\centering
	\includegraphics[width=0.5135\linewidth]{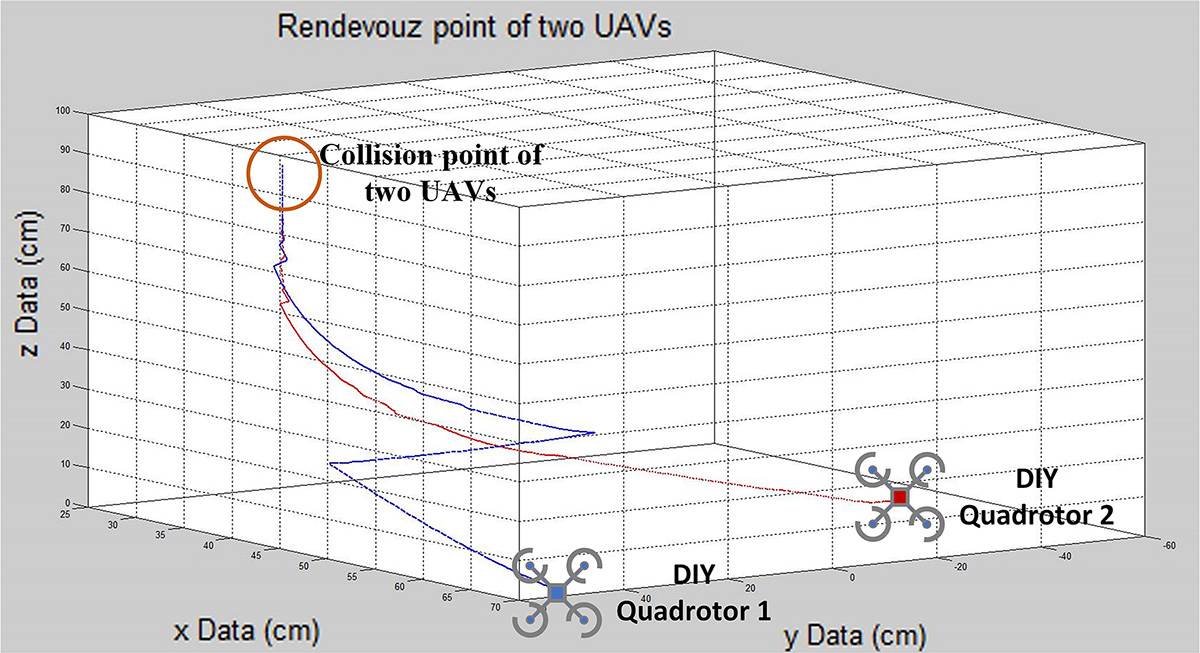}
	\caption{The position consensus between two UAVs via the original NI consensus protocol.}
	\label{fig:original_ni_consensus}
\end{figure}

This paper researches time-varying formation control approaches for multi-robot systems based on the Negative Imaginary (NI) consensus systems theory. Since it was first introduced in 2010 \cite{Pertersen}, NI systems theory is becoming popular in many high technology applications, especially in nano-positioning control \cite{Mabrok}, vibration control of flexible structures \cite{Abdullahi} and multi-DOF robotic arms. According to \cite{Wang1} and \cite{Wang2}, closed-loop multi-systems can achieve high robustness with respect to external disturbances when certain conditions of NI theory are met in the plant and controllers. However, in practice, the cooperative architecture of this theory shows four disadvantages, which is proven by our simulation as shown in Fig. \ref{fig:original_ni_consensus}: (1) the NI-consensus-based formation control architecture is not developed in this literature, (2) all agents are controlled to reach one unique rendezvous point, resulting in position consensus at the same point which would lead to collision, (3) the asymmetrical consensus equations between master and slave can cause practical difficulties in tuning the gain values of the SNI controllers, and (4) the input setpoints of each agent such as position/velocity set points are not mentioned in the overall network plant of this theory. To handle these problems, we have previously proposed and tested a new NI architecture for multi-UAV operation \cite{Tran17}. Nevertheless, our previous paper presented only the experimental results of a homogeneous UAV system with an invariant formation and did not mention any potential application for the NI formation control protocol.

Therefore, the significant contributions of this paper compared to our earlier results are four folds: (1) providing an improved formation control protocol via a more effective strategy using next-position prediction of the master and relative distance between agents; (2) simultaneously guaranteeing both stability of distributed time-variant formation and obstacle avoidance control, relative distance modification and velocity control loops via the NI stability criteria; (3) developing a distributed time-varying formation controller for multi-vehicles system using NI systems theory; (4) presenting a new obstacle avoidance approach by varying the formation shape; and (5) considering necessary and sufficient conditions for all obstacle avoidance situations and proposing effective solutions. An additional contribution of our work is introducing a practical strategy for UAV-UGV coordination since not many recent studies consider a consensus methodology for these two different types of unmanned robots. 

In order to test our new methodology, three indoor experiments implementing our enhanced NI-systems cooperative approach are conducted using both UAV and UGV platforms. The first test is carried out using two UGVs and one UAV. The autonomous vehicles must maintain a small invariant triangular pattern while the UGV (master) follows a pre-defined rectangular trajectory. The purpose of this test is to prove the effectiveness and stabilization of our proposed method when two types of robots coordinate to perform an indoor exploration task. The second test, which is more complicated than the first one, is used to show the obstacle avoidance capability of our varying formation method in case the master (UAV) predicts a collision with an obstacle and the master (UGV) predicts a collision with multiple obstacles on either side of its desired path. The final test involves three UGVs with randomly located obstacles.

Moreover, in outdoor cooperative scenarios, where the UAV and UGVs are tasked to explore uncertain environment, situations may occur when an absolute position sensor (like GPS) is only available to one vehicle at a time. Because the absolute position navigation may not be available to all vehicles (e.g. GPS interference caused by urban canyon effect), we assume the other vehicles are only able to measure their relative distance using sensors such as machine vision sensors or radio beacons. Therefore, the relative distances between the autonomous robots are used as key components in our cooperative method during indoor experiments. This information is simulated and measured by a Vicon Motion Capture system. Finally, due to the sharp interior angle vertexes of a rectangle (generally being 90 degree), yaw angle consensus among the UGVs is also considered in this paper. This additional angle consensus allows the UGVs to perform smoother movement following a trajectory with sharp corners like a rectangle by rotating simultaneously around the vertical (Z) axis, especially when the master UGV reaches a vertex of the rectangle.

The rest of this paper is organized as follows: Section II introduces the hardware/software configuration and architecture design for the formation experiment. Preliminary graph theory and the NI theory is reviewed in Section III. The dynamic models of the UAV and UGV are analyzed in Section IV. The formation tracking control architecture for an autonomous team (UAVs and UGVs) is studied in Section V. Stability of the whole system is verified in Section VI. Simulation results from implementing the NI formation control architecture in a team of multi-vehicles are shown in Section VII. The geometrical approach for obstacle avoidance is illustrated in Section VIII. Finally, experimental results are presented in Section IX, followed by conclusions drawn in Section X.

Throughout this paper, for simplicity of notation, we let \textit{1$_{n}$} represent a column vector of size \textit{n} with all elements being 1. Let \textit{$\otimes$} represent the Kronecker product and \textit{[P(s), P$_{s}$(s)]} denote the positive feedback interconnection between transfer functions \textit{P(s)} and \textit{P$_{s}$(s)}.

\section{Experimental Apparatus}
\subsection{Hardware/Software Configuration}

One UAV (F450 quadrotor platform), three UGVs (Pioneer P3-AT and P3-DX) and four obstacles (plastic packing boxes) are used to conduct our experiments. The dimension of each packing box is \textit{36cmW$\times$36cmD$\times$36cmH}. The size of each UAV or UGV is approximately \textit{30cmW$\times$30cmD}. The position and orientation of each robot are broadcasted continuously from a ground station (GS) where the movements of the UAV and UGVs in three dimensions are analyzed by the Vicon Motion Capture System software that is installed in the station facilities. The master UAV/UGV utilizes this information to correct its trajectory relative to the next waypoint while the two slaves (UGVs or UAV) predict the next position of the master from its estimated velocity and recalculate the relative distance between the master and the slave before adjusting its position to preserve the determined formation in real time. The information exchange protocol between the GS and the robots is achieved using a WiFi network and the Robot Operating System (ROS) framework.
	
\subsection{Overall Architecture}

\begin{figure}
	\centering
	\includegraphics[width=0.75\linewidth]{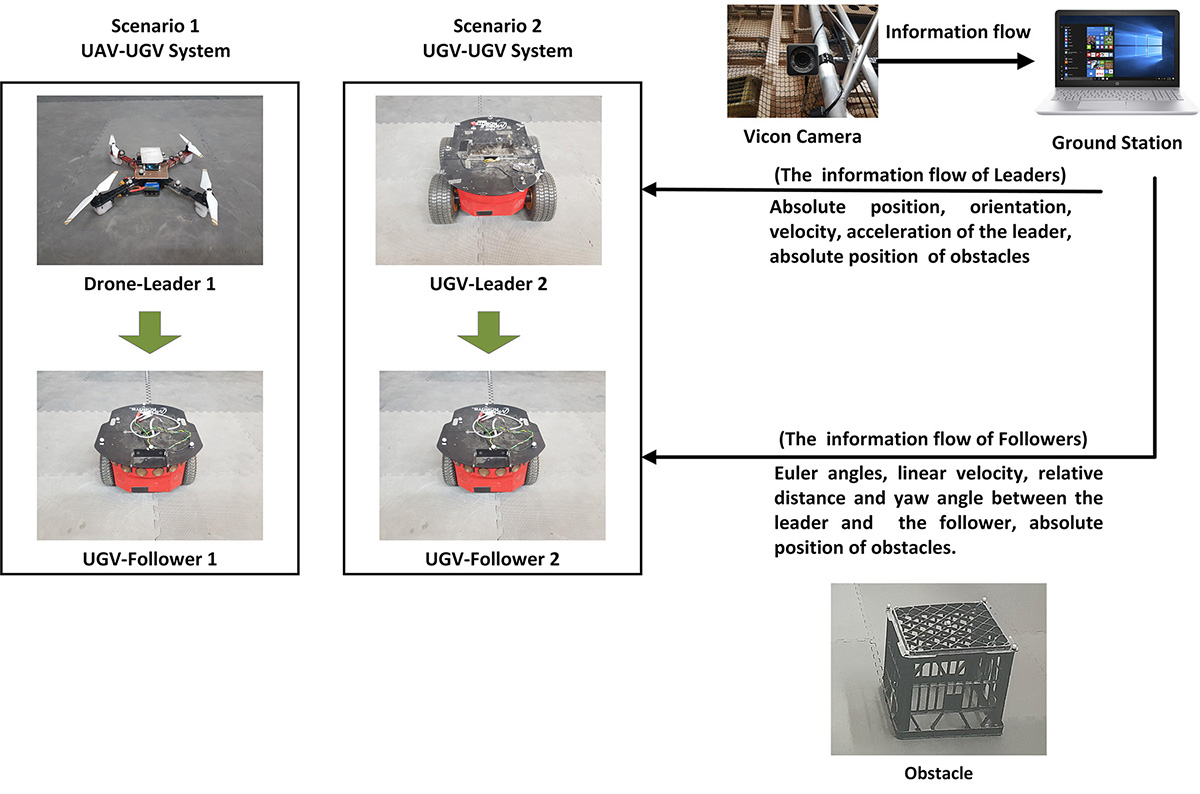}
	\caption{Overall architecture of the UAV-UGV and UGV-UGV formation.}
	\label{fig:architecture}
\end{figure}

Our control architecture is separated into two layers (Fig. \ref{fig:architecture}): UGV-Leader layer and Leader-Follower layer. The functionality and the control design of each layer will be described in the next sections.

\section{Preliminaries of graph theory and Negative-Imaginary theory}

\subsection{The Basics of Graph Theory}

Graph theory can be used to effectively describe the inter-agent communication including N unmanned robots. Each robot is represented as a vertex. Let \textit{G} = \textit{($\vartheta$,$\varepsilon$)} denote a directed graph with \textit{N} vertexes and \textit{L} edges (\textit{N$\times$L} matrix), such that \textit{$\vartheta$} = \textit{(v$_{1}$, v$_{2}$,..., v$_{N}$)} and  \textit{$\varepsilon$ $\subseteq$$\{$(v$_{i}$,v$_{j}$): v$_{i}$,v$_{j}$ $\in$ $\vartheta$, v$_{i}$ $\neq$ v$_{j}$$\}$} mathematically describe the finite vertexes set and the ordered edges set, respectively. Suppose that each edge of \textit{G} is assigned an orientation, which is arbitrary but fixed. The (vertex-edge) incidence matrix of \textit{G}, denoted by \textit{Q(i)}, is a \textit{N$\times$L} matrix defined as follows:

\begin{equation}
Q_{i}=% 
\begin{cases}
q_{ve} = 1 &\text{if $v_{i}$ is the head of edge $e_{v_{i},v_{j}}$}, \\
q_{ve} = -1 &\text{if $v_{i}$ is the tail of edge $e_{v_{i},v_{j}}$},\\
q_{ve} = 0 &\text{if $v_{i}$ is not connected to $e_{v_{i},v_{j}}$}.
\end{cases}
\end{equation}

Note that the incidence matrix has a column sum equal to zero, which is assumed from the fact that every edge must have exactly one head and one tail.
Extending this graph theory, given an arbitrary oriented set of the edge \textit{$\varepsilon$}, the graph Laplacian, \textit{$L(G)$}, is defined as \textit{L(G)= Q(G)Q(G)$^{T}$}.

For convenience in interpreting further the UAV-UGV cooperative theory, some mathematical information graph matrices are newly described.

\subsubsection{The consensus matrix} Let the \textit{N$\times$L} matrix  \textit{Q$_{c}$} represent a consensus matrix which corresponds to a graph \textit{($\vartheta$$_{c}$,$\varepsilon$$_{c}$)} that has \textit{N} vertexes and \textit{L} is the number of communication orientations. This \textit{N$\times$L} matrix indicates which robot at vertex v$_{i}$ in $\vartheta$$_{c}$ will send its consensus information to the adjacent vertexes v$_{j}$ in $\varepsilon$$_{c}$.

\begin{equation}
Q_{c}=% 
\begin{cases}
C_{ij} = 0 &\text{if \textit{Robot$_{i}$} is the head of $e_{v_{i},v_{j}}$} \\ &\text{for consensus}, \\
C_{ij} = -1 &\text{if \textit{Robot$_{i}$} is the tail of $e_{v_{i},v_{j}}$}\\ &\text{for consensus},\\
C_{ij} = 0 &\text{otherwise}.
\end{cases}
\end{equation}

\subsubsection{The reference matrix}

The (N$\times$L) reference matrix \textit{Q$_{r}$ = diag$_{i=1}^{n}${R$_{i}$}} describes which node or UAV in the group will directly follow to the reference trajectories: 

\begin{equation}
Q_{r}=% 
\begin{cases}
R_{i} = 1 &\text{if \textit{Robot$_{i}$} directly follows the reference}
\\ &\text{path}, \\
R_{i} = 0 &\text{if \textit{Robot$_{i}$} does not directly follow the} 
\\ &\text{reference path}. \\
\end{cases}
\end{equation}

\subsection{NI Systems Theory}

The definition of an SNI system was firstly introduced in \cite{Pertersen}. According to Lemma \ref{SNIsystem}, its Nyquist plot should start from an arbitrary point on the real axis, curve below this axis and end at the center point (0,0). Similarly, with a less strict criteria as compared to an SNI system, the definition of an NI system is presented in Lemma \ref{NIsystem}.

\begin{lemma}[Petersen, I.R. and Lanzon, A.,2010]\label{SNIsystem} 
For a single input/single output (SISO) case, a transfer function \textit{P(s)} is called strictly negative imaginary (SNI) if its Nyquist plot is located strictly below the real axis for all positive frequencies in \textit{s} domain; i.e, \textit{P(s)} $\in$ \textit{SNI} if \textit{j(P(j$\omega$)-P$^{*}$(j$\omega$))$>$0} for all \textit{$\omega$$>$0}.  
\end{lemma}

\begin{lemma}[Petersen, I.R. and Lanzon, A.,2010]\label{NIsystem} 
\textit{P(s)} is negative imaginary (NI) if its Nyquist plot for $\omega$ $\geqslant$ \textit{0} is contained in the lower half of the complex plane. This Nyquist plot can touch, but not cross the real axis; i.e, \textit{P(s)} $\in$ \textit{NI} if \textit{{j(P(j$\omega$)-P$^{*}$(j$\omega$))$\geqslant$0}}
\end{lemma}

when a SNI/NI plant system is achieved, the internal stability of a positive feedback interconnection between this plant and its NI/SNI controller is illustrated as in Lemma \ref{NI stability}.

\begin{lemma}[Petersen, I.R. and Lanzon, A.,2010]\label{NI stability} 
Considering the case in which an SNI/NI plant \textit{M(s)} and an NI/SNI controller \textit{N(s)} such that M($\infty$)N($\infty$) = 0 and N($\infty$) ≥ 0 are interconnected by a positive feedback, it follows that the corresponding loop transfer function \textit{L(s) = M(s)N(s)} is internally stable if the DC gain condition at zero frequency \textit{M(0)N(0) $<$ 1} is satisfied.
\end{lemma}

\begin{remark}
Since the \textit{(-1,0)} critical point is never enclosed by the Nyquist plot of a SNI or NI system, it is assumed that any SNI/NI system has a robust stability. 
\end{remark}

The major concept of this theory can be extended further to MIMO and cooperative control applications \cite{Wang1} and \cite{Wang2}. Given a group of SNI/NI systems or a multi-input multi-output (MIMO) system, the collaboration of SNI/NI agents is achieved when these systems are connected to their SNI controllers via a cooperative information graph. The SNI/NI-agents cooperation is formulated as follows:

\begin{equation}
\bar{y} = \bar{P}(s)\bar{u} = (Q_{i}^{T}\otimes{I_{n}})diag_{i=1}^{n}{P_{i}(s)}(Q_{i}\otimes{I_{n}})\bar{u}
\end{equation}
\begin{equation}
\bar{u} = \bar{P}_{s}(s)\bar{y}
\end{equation}

where $\overline{u}\in\mathbb{R}^{ln\times1}$ and $\overline{y}\in\mathbb{R}^{ln\times1}$ are the input and output of the overall network plant, respectively.  $\bar{P}_{s}(s)\in\mathbb{R}^{ln\times ln}$ is the group of SNI/NI consensus controllers for the multiple NI/SNI systems, indicated as $P_{i}(s)\in\mathbb{R}^{m_{i}\times 1}$. Thus \textit{P}$_{i}(s)$ is a column vector of size \textit{M} which contains a set of \textit{m}$_{i}$ coordinated outputs. 

\begin{figure}
	\centering
	\includegraphics[width=0.75\linewidth]{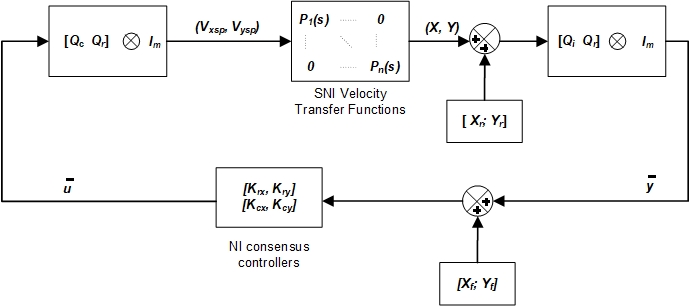}
	\caption{The origin NI-system consensus architecture.}
	\label{fig:old_structure}
\end{figure}

Based on the consensus theory of Wang \cite{Wang1,Wang2}, an NI-systems consensus architecture for the multi-UAV systems was proposed in \cite{Tran17}. A reference matrix, consensus matrix as well as an offset distance between the UAVs are added to the original structure to solve existing formation and consensus problems as shown in Fig. \ref{fig:old_structure}. The overall formulas for the whole cooperative structure are given as:

\begin{equation}
X_{r} = 1_{n}\otimes{r},
\end{equation}
\begin{equation}
e_{i} = y_{i} + X_{r},
\end{equation}
\begin{equation}
\bar{y} = ([Q_{i}~Q_{r}]\otimes{I_{m}})e_{i},
\end{equation}
\begin{equation}
e_{f} = \bar{y} + X_{f},
\end{equation}
\begin{equation}
u= ([Q_{c}~Q_{r}]\otimes{I_{m}})diag_{i=1}^{n}{P_{s}(s)}e_{f}
\end{equation}

where \textit{r} $\in\mathbb{R}^{2\times1}$ denotes the reference position on the plane of the master, \textit{X$_{r}$} $\in\mathbb{R}^{2n\times1}$ is the reference matrix, and \textit{x}$_{f}$ corresponds to the relative position between the slave UAVs and the master UAV in the configured formation. \textit{u} $\in\mathbb{R}^{m\times{1}}$ is the velocity set point input of each UAV on x and y axes while the output \textit{y}  $\in\mathbb{R}^{m \times{1}}$ is the current position of each UAV. $\overline{u}\in\mathbb{R}^{lm\times1}$ and $\overline{y}\in\mathbb{R}^{lm\times1}$ are the input and output of overall network plant. \textit{e$_{r}$} is the error between the desired position of the master and its current one, while \textit{e$_{f}$} is the error between the desired relative position and the current one among the UAVs.  $P_{s}(s)\in\mathbb{R}^{lm\times lm}$ is the group of SNI consensus and tracking controllers for the group of UAVs. 

The simple formation experiment conducted in this paper shows that consensus based on NI-systems cooperative control between UAVs is guaranteed, but the position response of the remaining UAVs is somewhat slow. This small delay in the position response causes an unsatisfactory formation shape which is also discussed in the conclusion of the Section VII.

In an effort to improve the UAV-UGV formation performance based on the consensus algorithm, we will outline an innovative architecture in the next sections.

\section{Dynamic Modeling for Velocity Control of The UAV and The UGV}

\begin{figure}
	\centering
	\includegraphics[width=0.75\linewidth]{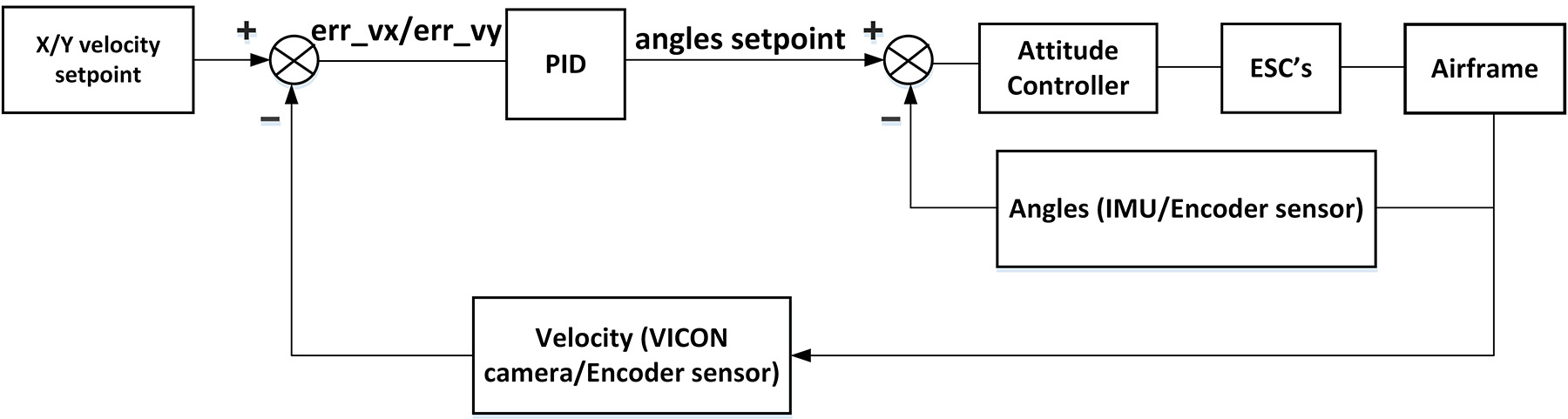}
	\caption{A PID controller structure for controlling the velocity of robots.}
	\label{fig:x_y_velocity}
\end{figure}

First, the linear velocities of the UAV/UGV in the x and y axes are stabilized by using a PID controller, which is tuned to satisfy the closed-loop SNI system requirement. The proposed velocity control structure is illustrated in Fig. \ref{fig:x_y_velocity}.

During experimental tests, all relevant data of the master and the slaves (position, linear velocity, acceleration, Euler angles) were automatically collected by the ground station to serve the purpose of finding the closed-loop velocity transfer functions in the lateral and longitudinal directions. In the next step, this data was analyzed using MATLAB to determine the closed-loop transfer functions between the velocity set point input (\textit{velxsp/velysp}) and the absolute position output (\textit{posx/posy}) for the UAV/UGV on the x and y axes (Table \ref{UGVvelocitytransferfunction}, \ref{UAVvelocitytransferfunction}). We use an ARMAX model to represent the performance of these transfer functions. The overall dynamic equations for the UAV and the UGV in the z domain have a general form as follows:

\begin{equation}
A(z)y(k)=B(z)u(k-n)+e(k)
\end{equation}

where \textit{u(k)} is the system input, \textit{y(k)} is the system output, \textit{n} is the system delay, \textit{k} is the present time and \textit{e(k)} is the disturbance in the system.

\begin{table}
	\begin{center}
		\caption{The UGV transfer functions between velocity x/velocity y set point and actual position x/ position y}
		\label{UGVvelocitytransferfunction}
		\begin{tabular}{|c|c|}
			\hline 
			& Transfer function \\ 
			\hline 
			Velx & velxsp/posx=$\frac{-0.043s^{3}+5.16s^{2}-1860.11s+54489.05}{s^{4}+86.155s^{3}+54490.53s^{2}+29510.63s+1481.6}$\\ 
			\hline 
			Vely & velysp/posy=$\frac{-0.043s^{3}+4.24s^{2}-1494.74s+48347.83}{s^{4}+88.9s^{3}+54883.91s^{2}+49242.03s+1266.25}$ \\ 
			\hline 
		\end{tabular} 
	\end{center}
\end{table}

\begin{table}
	\begin{center}
		\caption{The UAV transfer functions between velocity x/velocity y set point and actual position x/ position y}
		\label{UAVvelocitytransferfunction}
		\begin{tabular}{|c|c|}
			\hline 
			& Transfer function \\ 
			\hline 
			Velx&velxsp/posx=$\frac{-0.0426 s^{3}+6.76s^{2}-153.90s+33206.86}{s^{4}+86.15s^{3}+54490.53s^{2}+24730.20s+1161.87}$\\ 
			\hline 
			Vely&velysp/posy=$\frac{-0.05 s^{3}+-0.52s^{2}-1144.04s+29990.15}{s^{4}+101.06s^{3}+52978.06s^{2}+23472.45s+2291.96}$ \\ 
			\hline 
		\end{tabular} 
	\end{center}
\end{table}

\begin{figure}
	\centering
	\includegraphics[width=0.75\linewidth]{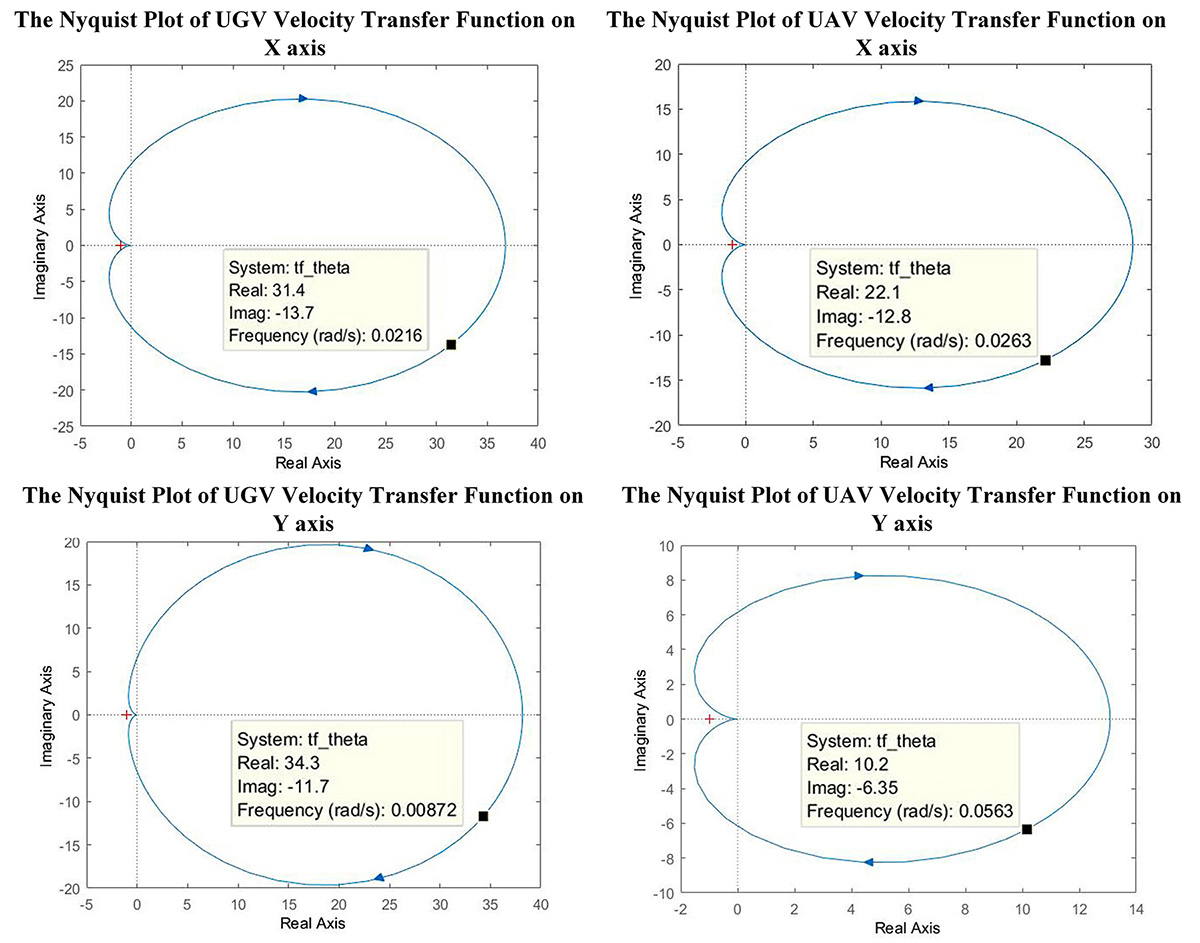}
	\caption{The Nyquist plots of the UGV and UAV velocity transfer functions for the \textit{x} and \textit{y} axes.}
	\label{fig:Robotnyquistplot}
\end{figure}

Based on the transfer functions found, the Nyquist diagrams of the UAV and UGVs are plotted to identify if the transfer functions are SNI. As seen in Fig. \ref{fig:Robotnyquistplot}, the linear velocity transfer functions for both the UAV and the UGVs satisfy the SNI conditions. 

Similarly, the yaw rate transfer functions for UGVs are also shown to be SNI systems using their Nyquist plots.

\section{Architecture design}

\begin{figure}
	\centering
	\includegraphics[width=0.4\linewidth]{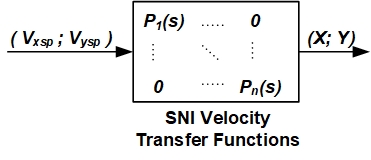}
	\caption{Multiple SNI plants.}
	\label{fig:SNItrans}
\end{figure}

\begin{figure}
	\centering
	\includegraphics[width=0.88\linewidth]{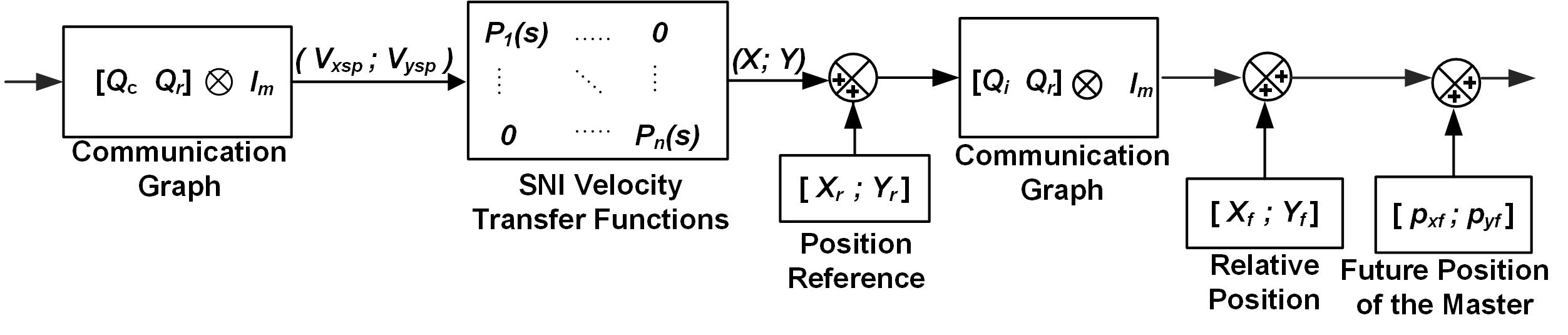}
	\caption{Overall network plant.}
	\label{fig:overall_plant}
\end{figure}

In this section, an architecture for the NI-system formation and consensus control is newly designed. Firstly, all of the SNI velocity transfer functions for the x and y axes, found in the previous section, are combined to form an SNI plant. The inputs of this plant are the velocity set points of the SNI systems \textit{(V$_{xsp}$, V$_{ysp}$)} while its outputs correspond to the current locations of the vehicles on the x and y axes as shown in Fig. \ref{fig:SNItrans}. In the next step, the network matrix \textit{([Q$_{i}$~Q$_{r}$]$\otimes${I$_{m}$})} is appended to the output of plant in order to calculate the relative distances between the robots and the position error between the master and the slaves \textit{(dx;dy;dx$_{1r}$;dy$_{1r}$)}. Similarly, the matrix \textit{([Q$_{c}$~Q$_{r}$]$\otimes${I$_{m}$})} is appended to the input of the plant to determine the desired velocity values for each robot via the NI consensus controller gains \textit{K}, the desired relative distance between robots \textit{(X$_{f}$, Y$_{f}$)} and the next position prediction of the master \textit{(p$_{xf}$, p$_{yf}$)}. This control architecture is referred to as the overall network plant for consensus (Fig. \ref{fig:overall_plant}).

\begin{figure}
	\centering
	\includegraphics[width=0.75\linewidth]{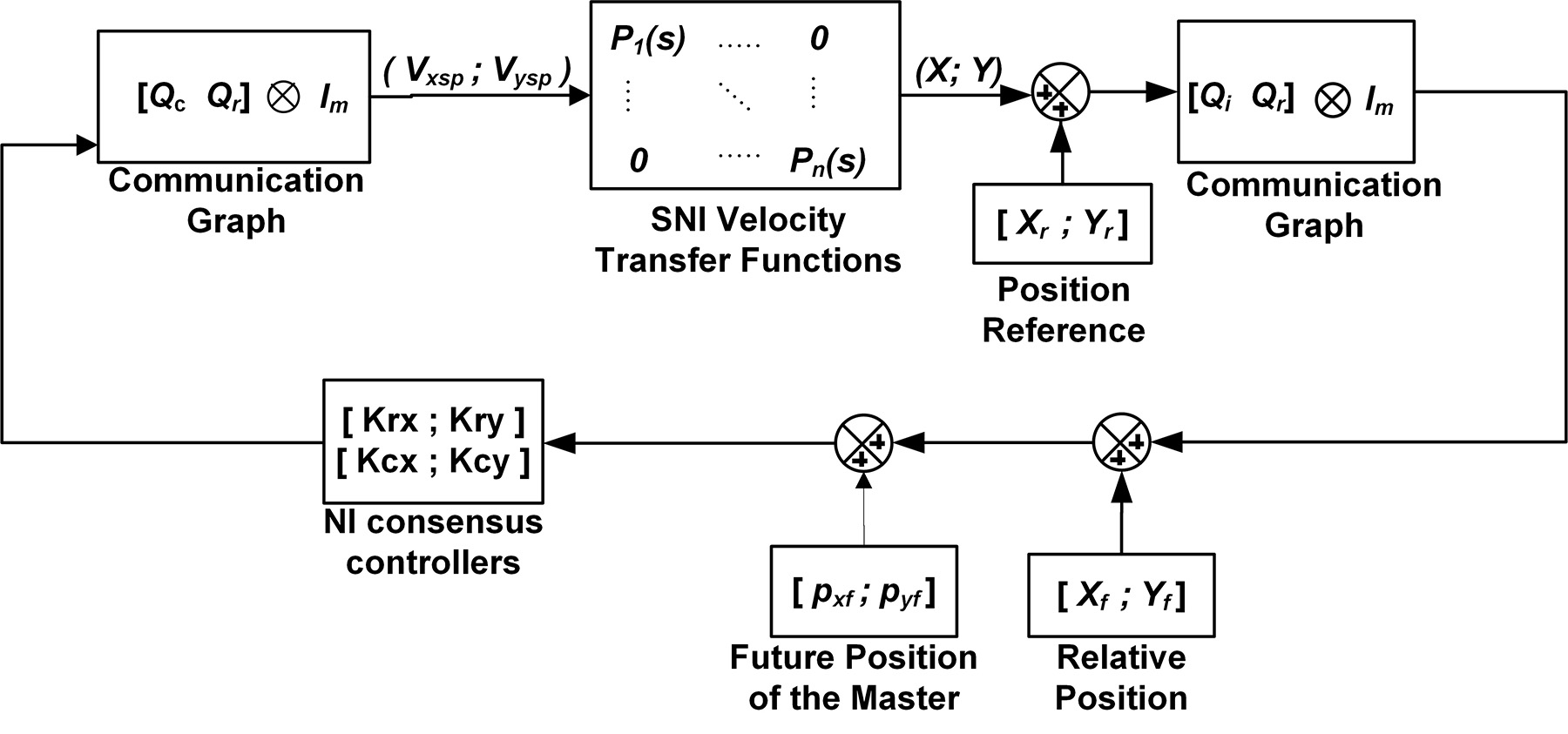}
	\caption{The consensus and formation control architecture for SNI multi-vehicle systems with a corresponding communication graph.}
	\label{fig:overall_structure}
\end{figure}

The distance between the current position of the master and its next position \textit{(p$_{xf}$, p$_{yf}$)} in Fig. \ref{fig:overall_structure} is predicted from its current velocity on the horizontal plane. The corresponding equations are as follows:

\begin{equation}
\begin{bmatrix}
p_{xf}\\
p_{yf}
\end{bmatrix}
=\begin{bmatrix}
~\text{V$_{xm}$(dt)}\\
~\text{V$_{ym}$(dt)}
\end{bmatrix}
\end{equation}

where \textit{V$_{xm}$} and \textit{V$_{ym}$} are the master horizontal and vertical velocities, respectively. This data is sent continuously to every slave agent. Here, \textit{dt} represents the sample time for the whole system.

Finally, the overall network plant is then combined with the remaining parts of the consensus controller, including the NI controllers, the relative positions and the future position prediction, to create a new formation consensus control method as shown in Fig. \ref{fig:overall_structure}.

\begin{figure}
	\centering
	\includegraphics[width=0.65\linewidth]{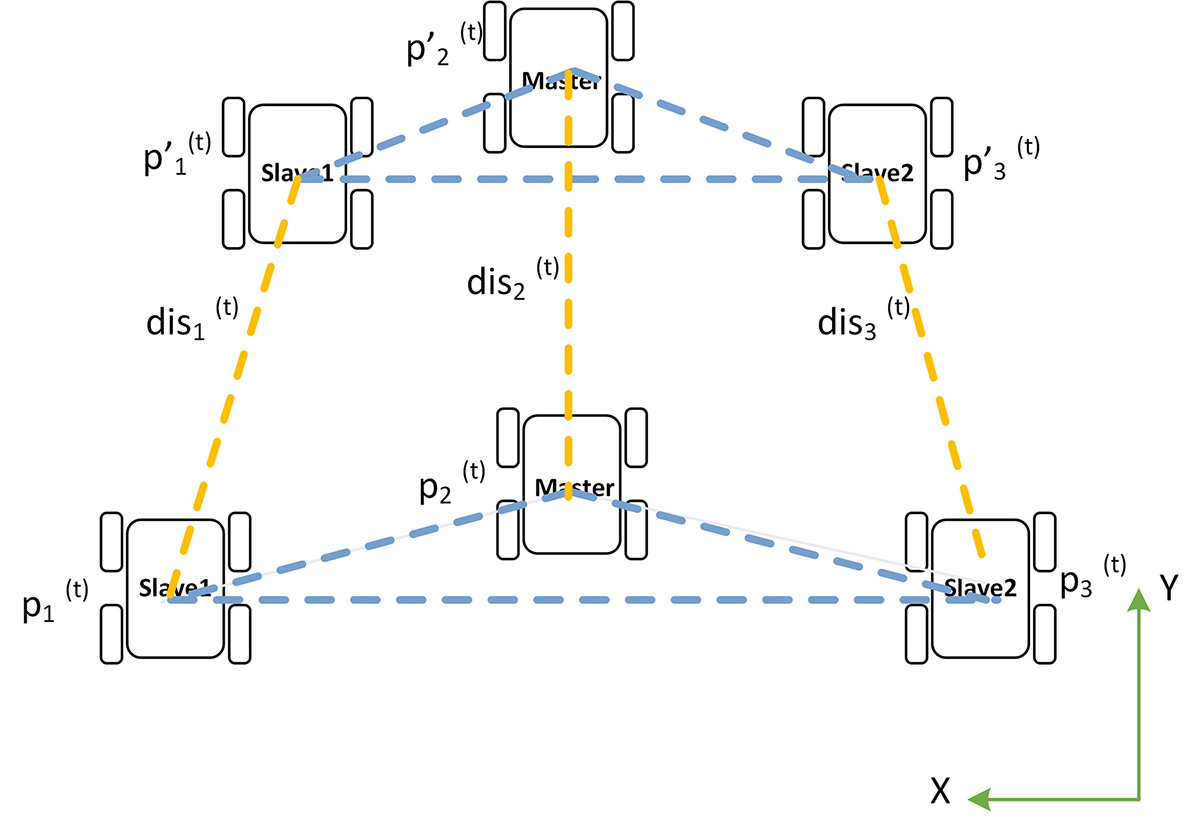}
	\caption{Time-varying formation of three robots}
	\label{fig:time-varying}
\end{figure}

Based on the desired distance and traveling time between the former and latter formation, the velocity setpoints sent to each robot can be formulated as shown in Equation \ref{eq:2} and Fig. \ref{fig:time-varying}. 

\begin{equation}\label{eq:2}
\sum_{i=1}^n (Vxsp_{i},Vysp_{i})=\sum_{i=1}^n (disx_{i}^{(t)}/t,disy_{i}^{(t)}/t)
\end{equation}
where \textit{n} is the number of agents participating in formation transformation. \textit{$(disx_{i}^{(t)}, disy_{i}^{(t)})$} are the relative distances between the old and new formation on the x and y axes. \textit{t} is the desired time interval to reach a new formation.

As a result, the consensus and reference gain should be adaptive to obtain the given velocity.

\begin{theorem}
Multi-agent systems can obtain a time-varying formation tracking on the x-y plane if for any given bounded initial states, the following condition holds:
\begin{equation}\label{1}
\lim_{t \to dest} (\sum_{i=1}^n \acute{p}_{i}^{(t)} - dis_{i}^{(t)}  - p_{i}^{(t)} ) = 0_{2\times 1}
\end{equation}
\end{theorem}
where \textit{dest} is the expected traveling time to achieve a new formation. \textit{p$_{i}^{(t)}$, $\acute{p}_{i}^{(t)}$} are expressed as the current and desired position of each robot in the varying formation. 

In the case where $\lim_{t \to dest} \sum_{i=1}^{n} dis_{i}^{(t)}$  = $0_{2\times1}$, it is assumed from Eq. \eqref{1} that $\lim_{t \to dest} (\sum_{i=1}^n \acute{p}_{i}^{(t)}$ - $p_{i}^{(t)}) = 0_{2\times1}$. Therefore, this theorem becomes the definition of targets pursuing problems intended for multi-agent systems.

Based on this new strategy, the multi-vehicles under the control of the adaptive NI controllers and the information network topology will be guaranteed not only to maintain a desired spatial pattern but also successfully modify their formation over time.

\section{Stability proof of the enhanced NI formation control architecture}

\begin{figure}
	\centering
	\includegraphics[width=0.7\linewidth]{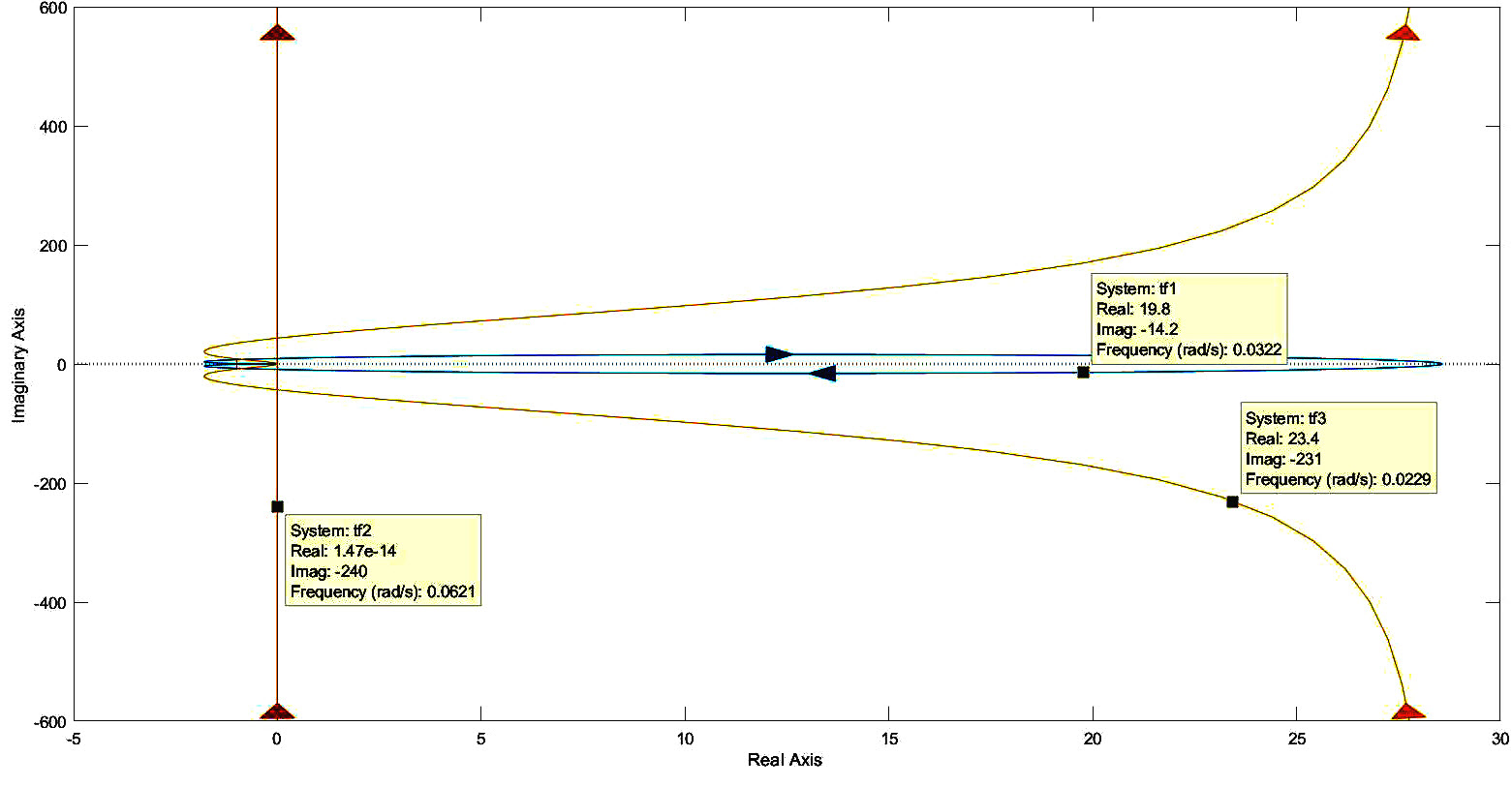}
	\caption{SNI property of a positive connection between an SNI system and an NI system}
	\label{fig:SNI_system1}
\end{figure}

We assume that \textit{M(s)} and \textit{N(s)} are the SNI velocity transfer function and its NI controller for one slave UAV on the x axis respectively. As presented in Table \ref{UAVvelocitytransferfunction} and Fig. \ref{fig:Robotnyquistplot}, the \textit{M(s)} transfer function (TF1) is an SNI system. Unfortunately, the position prediction term is only an NI system (TF2) as shown in Lemma \ref{NIsystem}. It is outlined in Lemma \ref{SNI} that the whole structure is still SNI if two systems are linked together by a positive connection. As presented in the Fig. \ref{fig:overall_plant}, we may conclude that the complete plant (TF3) is SNI as shown in Fig. \ref{fig:SNI_system1}.

\begin{lemma}[Petersen, I.R. and Lanzon, A.,2010]\label{SNI} 
A positive connection between an NI plant and an SNI plant results in an SNI structure.
\end{lemma}

\begin{lemma}[Ghallab, et al., 2017]\label{Lyapunov} 
	Lyapunov-based stability proof verifies that the results given in Lemma \ref{NI stability} for the internal stability of feedback interconnections of NI systems is still correct.
\end{lemma}

\begin{figure}
	\centering
	\includegraphics[width=0.7\linewidth]{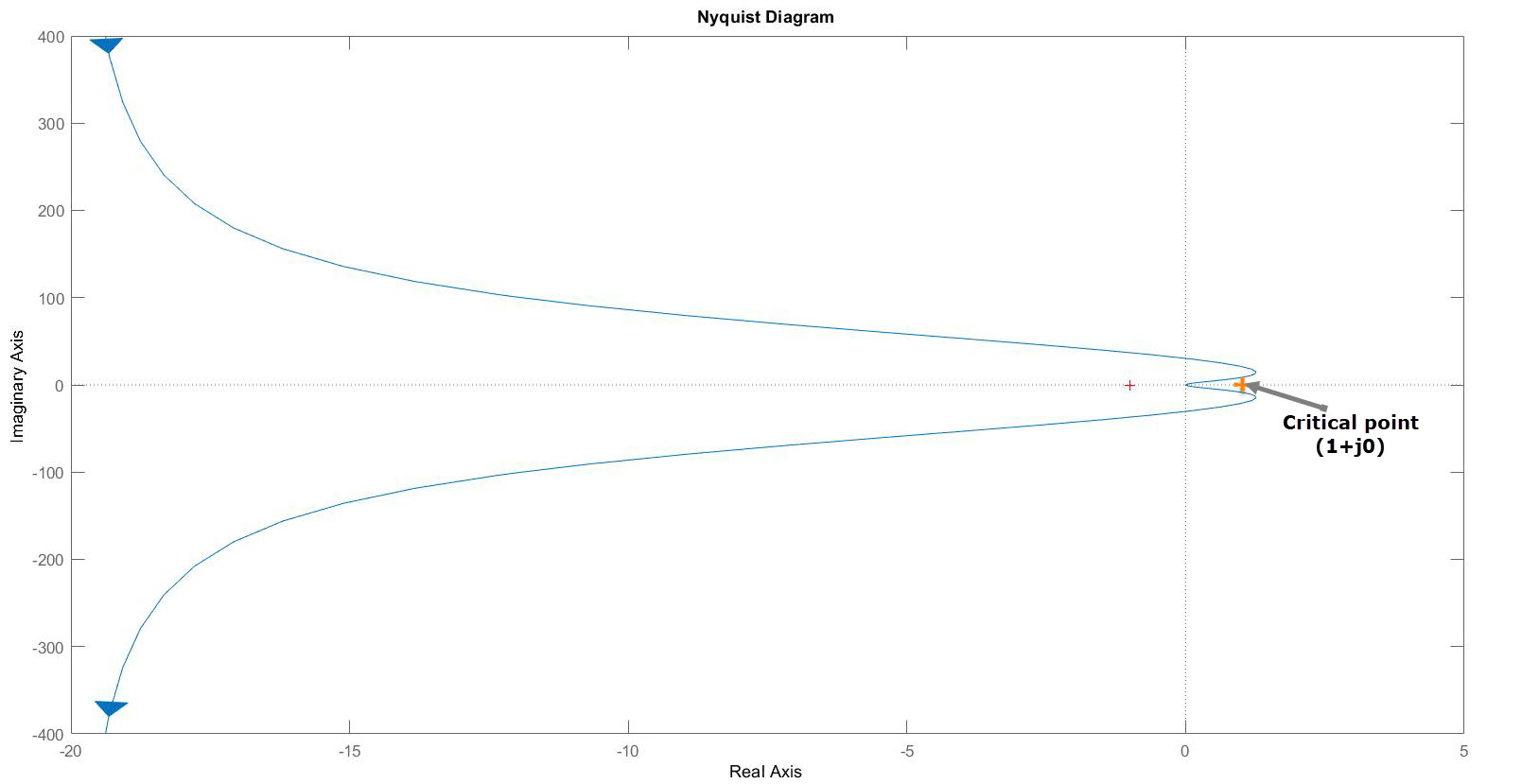}
	\caption{The stability result of the NI formation control protocol}
	\label{fig:prove2}
\end{figure}

Furthermore, it is mentioned in Lemma \ref{NI stability} that the condition of internal stability for the whole structure with the positive feedback interconnection is \textit{M(0)N(0) $<$ 1}. The DC gain of \textit{M(0)} equals to $\frac{33206.861}{1161.872}$ $\approx$ 28.58 while that of \textit{N(s)} is a constant gain of \textit{-0.7} as presented in Test 2. It is seen that the DC gain condition of \textit{M(s)}\textit{N(s)} is guaranteed to be less than 1 due to the negative constant gain value in \textit{N(s)} controller. As a result, \textit{M(s)} under the control of \textit{N(s)} is internally stable. Moreover, a Nyquist plot of whole architecture is drawn, and the result of stability is the same as in Lemma \ref{NI stability} since its figure never encircles the critical point \textit{(1$+$j0)} as shown in Fig. \ref{fig:prove2}.

Similarly, the stability of the Y axis SNI velocity transfer function and its NI controller as well as that of whole NI formation control structure are proved by the same approach.

\section{Simulation Results}

To validate the effectiveness of the proposed architecture and controllers, we simulated four consensus schemes. The NI controller used for the master UAV will receive reference points from the rectangular path planner and convert them into linear velocity set points. During the time the master navigates to each next waypoint, the slaves maintain its relative distance from the predicted future position of the master. Case 1 shows a complicated consensus and formation for the combination of one UAV and two UGVs with the time-varying formation while the yaw angle consensus is also presented in Case 2 to solve the position consensus problem at each vertex of the rectangular trajectory. Thanks to this approach, UGVs can make the position consensus better, especially at the sharp corner of two trajectories which has not been mentioned in recent studies \cite{Dong1} \cite{Tran17}. The overall equations describing the consensus and formation control for one master UAV and two slave UGVs, using the enhanced NI systems theory are summarized as follows:

\begin{equation}
Q_{i}=\begin{bmatrix}
1\qquad1\\
-1\qquad0\\
0\qquad-1
\end{bmatrix}
;
Q_{r}=\begin{bmatrix}
1\\
0\\
0
\end{bmatrix}
;
Q_{c}=\begin{bmatrix}
0\qquad0\\
-1\qquad0\\
-1\qquad0
\end{bmatrix}
\end{equation}

\begin{equation}
([Q_{i}~Q_{r}]\otimes{I_{2}}) = \begin{bmatrix}
1\qquad1\qquad1\\
-1\qquad0\qquad0\\
0\qquad-1\qquad0
\end{bmatrix}
\otimes
\begin{bmatrix}
1\qquad0\\
0\qquad1
\end{bmatrix}
\end{equation}

\begin{equation}
([Q_{c}~Q_{r}]\otimes{I_{2}}) = \begin{bmatrix}
0\qquad0\qquad1\\
-1\qquad0\qquad0\\
-1\qquad0\qquad0
\end{bmatrix}
\otimes
\begin{bmatrix}
1\qquad0\\
0\qquad1
\end{bmatrix}
\end{equation}

\begin{equation}
\begin{split}
u=\begin{bmatrix}
V_{xsp_{1}^{(t)}}\\
V_{ysp_{1}^{(t)}}\\
V_{xsp_{2}^{(t)}}\\
V_{ysp_{2}^{(t)}}\\
V_{xsp_{3}^{(t)}}\\
V_{ysp_{3}^{(t)}}
\end{bmatrix}
&=\begin{bmatrix}
~\text{Kr$_{x}$(X$_{r}$ + X$_{1}$)} \\
~\text{Kr$_{y}$(Y$_{r}$ + Y$_{1}$)} \\
~\text{Kc$_{x_{1}}$[X$_{f}$ + X$_{1}$ - X$_{2}$ + (Vx$_{1}$dt)]}\\
~\text{Kc$_{y_{1}}$[Y$_{f}$ + Y$_{1}$ - Y$_{2}$ + (Vy$_{1}$dt)]}\\
~\text{Kc$_{x_{2}}$[X$_{f}$ + X$_{1}$ - X$_{3}$ + (Vx$_{1}$dt)]}\\
~\text{Kc$_{y_{2}}$[Y$_{f}$ + Y$_{1}$ - Y$_{3}$ + (Vy$_{1}$dt)]}\\
\end{bmatrix}\\
&=\begin{bmatrix}
~\text{Kr$_{x}$(X$_{r}$ + X$_{1}$)} \\
~\text{Kr$_{y}$(Y$_{r}$ + Y$_{1}$)} \\
~\text{Kc$_{x_{1}}$[X$_{f}$ + d$_{x_{1}}$]}\\
~\text{Kc$_{y_{1}}$[Y$_{f}$ + d$_{y_{1}}$]}\\
~\text{Kc$_{x_{2}}$[X$_{f}$ + d$_{x_{2}}$]}\\
~\text{Kc$_{y_{2}}$[Y$_{f}$ + d$_{y_{2}}$]}
\end{bmatrix}
\end{split}
\end{equation}
\begin{equation}
k=\begin{bmatrix}
~\text{Kr$_{x}$}\\
~\text{Kr$_{y}$}\\
~\text{Kc$_{x_{1}}$}\\
~\text{Kc$_{y_{1}}$}\\
~\text{Kc$_{x_{2}}$}\\
~\text{Kc$_{y_{2}}$}
\end{bmatrix}
=\begin{bmatrix}
~\text{disx$_{1}^{(t)}$/t*(X$_{r}$ + X$_{1}$)} \\
~\text{disy$_{1}^{(t)}$/t*(Y$_{r}$ + Y$_{1}$)} \\
~\text{disx$_{x_{2}}^{(t)}$/t*[X$_{f}$ + d$_{x_{1}}$]}\\
~\text{disy$_{y_{2}}^{(t)}$/t*[Y$_{f}$ + d$_{y_{1}}$]}\\
~\text{disx$_{x_{3}}^{(t)}$/t*[X$_{f}$ + d$_{x_{2}}$]}\\
~\text{disy$_{y_{3}}^{(t)}$/t*[Y$_{f}$ + d$_{y_{2}}$]}
\end{bmatrix}
\end{equation}

In order to generate a rotation at the sharp corner for UGVs, the yaw angle of the follower is controlled to follow the yaw angle of the leader (Eq. \ref{eq:20}) instead of being determined by the position vectors on the x and y axes during the consensus, avoiding and returning processes (Eq. \ref{eq:21} and Eq. \ref{eq:22}). Meanwhile, the position consensus is completely turned off during the turns. 

\begin{equation}\label{eq:20}
\begin{split}
u=\begin{bmatrix}
\omega_{sp_{1}}\\
\omega_{sp_{2}}
\end{bmatrix}
&=\begin{bmatrix}
~\text{Kr$_{yaw}$($\Omega$$_{r}$ + $\Omega$$_{1}$)}\\
~\text{Kc$_{yaw}$[$\Omega$$_{f}$ + $\Omega$$_{1}$ - $\Omega$$_{2}$ + ($\omega$$_{1}$dt)]}
\end{bmatrix}\\
&=\begin{bmatrix}
~\text{Kr$_{yaw}$($\Omega$$_{r}$ + $\Omega$$_{1}$)} \\
~\text{Kc$_{yaw}$[$\Omega$$_{f}$ + $\Omega$$_{l}$ + ($\omega$$_{1}$dt)]}
\end{bmatrix}
\end{split}
\end{equation}

\begin{equation}\label{eq:21}
\omega_{sp_{m}} = atan2(Y_{sp_{m}}-Y_{m},X_{sp_{m}}-X_{m})
\end{equation}
\begin{equation}\label{eq:22}
\omega_{sp_{l}} = atan2(disY_{sp_{s}}-disY_{s},disX_{sp_{s}}-disX_{s})
\end{equation}

where \textit{Kr} and \textit{Kc} are the reference and consensus gain values of the NI controllers for velocity control on the x and y axes and the angular rate control around the z axis. Let \textit{$\omega$}$_{sp_{1}}$ and \textit{$\omega$}$_{sp_{2}}$ represent the angular rate set point of the UGVs. \textit{d}$_{x}$ and \textit{d}$_{y}$ are the relative positions between the leader and the follower on the x and y axes while \textit{$\Omega$}$_{l}$ is the relative angle between the leader UGV and the follower UGVs. 

The output equation of the SNI plant is:
\begin{equation}
y=\sum_{i=1}^{n} (X_{i}, Y_{i}, \omega_{i})
= P(s)u
\end{equation}

where \textit{P(s)} is the SNI velocity transfer function matrix of the two UAVs on the x and y axes. \textit{n} is the number of agents.

\begin{figure}
	\centering
	\includegraphics[width=0.85\linewidth]{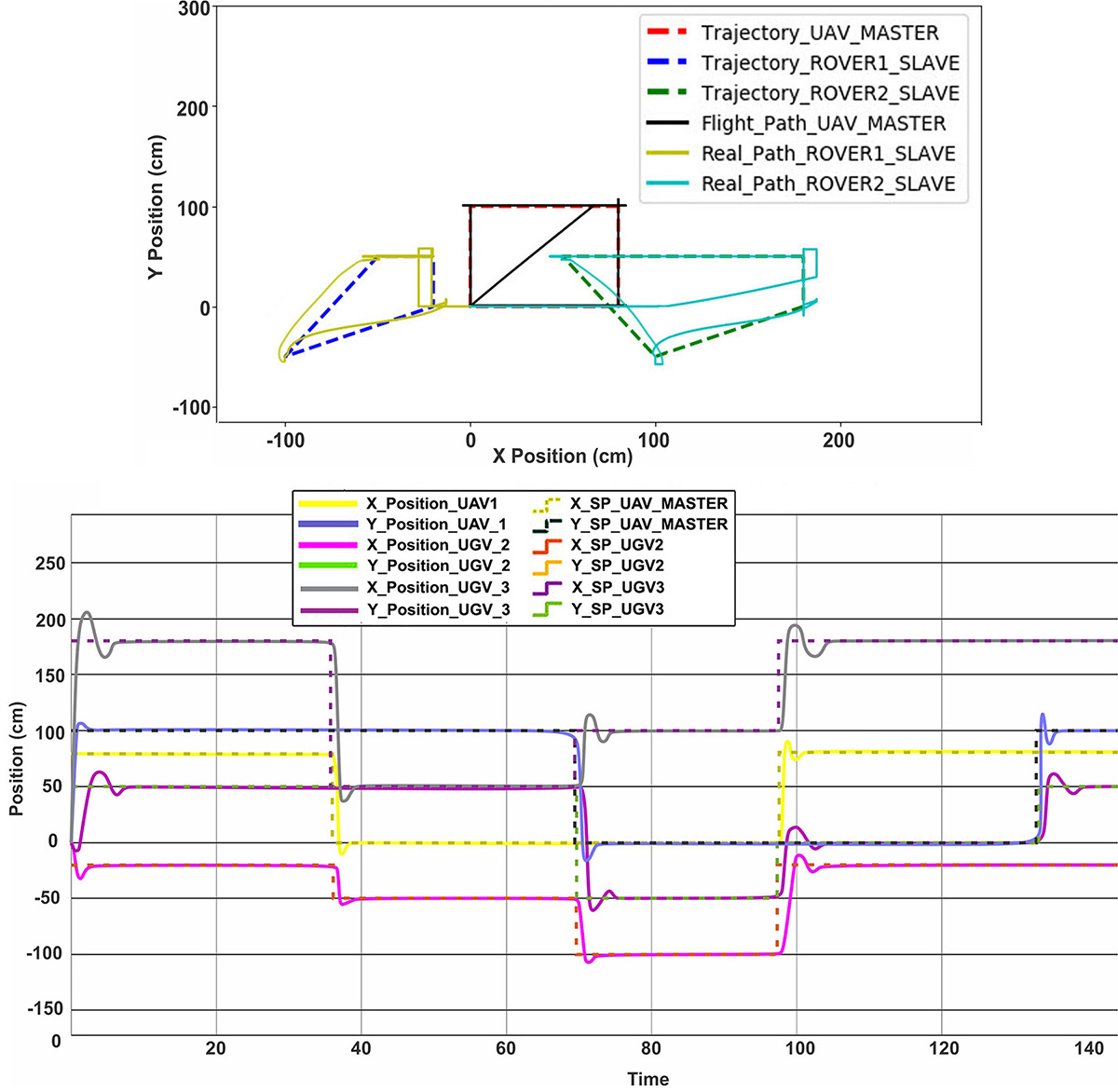}
	\caption{Simulation results obtained for a varied formation of the UAV-UGVs system on the plane for Case 1.}
	\label{fig:case1_simu}
\end{figure}

\begin{figure}
	\centering
	\includegraphics[width=0.18\linewidth]{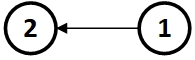}
	\caption{Communication Graph for Case1 and Case2.}
	\label{fig:Graph1}
\end{figure}

\textbf{Case 1}: The distances between robots on the x and y axes are varied at each vertex of the master's rectangular trajectory. \textit{X}$_{f}$ = [X$_{f_{1}}$, X$_{f_{2}}$]. \textit{X}$_{f_{1}}$ is chosen as [100,50], [50,50], [100,50], [100,0] and \textit{X}$_{f_{2}}$ is [-100,50], [-50,50], [-100,50], [-100,0]. The communication topology is given in Fig. \ref{fig:Graph1}. The \textit{X}$_{r}$ parameter will be assigned with the waypoints of the rectangular trajectory. The four vertexes of this rectangle in the x-y coordinate plane are [0, 100], [0, 0], [80, 0], [80, 100]. Desired time interval for formation variation is 2 seconds. Therefore, the initial values for NI controllers are selected as (-0.002, -0.5, -0.5, -0.5, -0.5, -0.5). As seen in Fig. \ref{fig:case1_simu}, \textit{UAV$_{1}$}, \textit{UGV$_{2}$} and \textit{UGV$_{3}$} well maintains the varied formation over time under the control of NI controllers. The UGVs obtain the newly desired relative distance in approximately 2 seconds.

\begin{figure}
	\centering
	\includegraphics[width=0.6\linewidth]{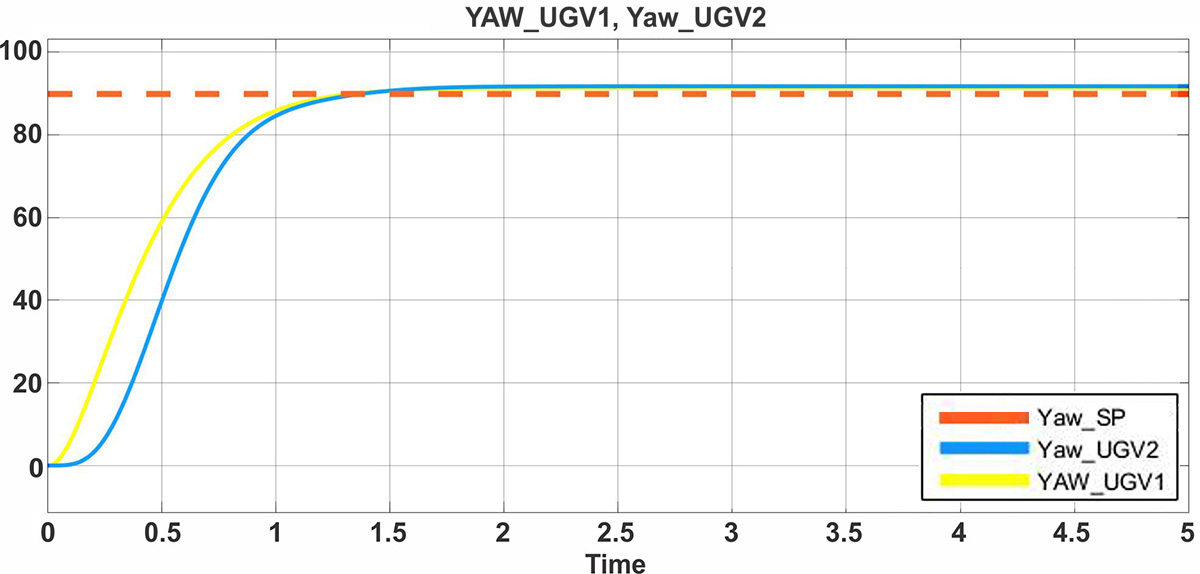}
	\caption{Simulation results for a yaw angle consensus involving the master UGV and the slave UGV.}
	\label{fig:case2_yaw_angle}
\end{figure}

\textbf{Case 2}: The communication topology is given in Fig. \ref{fig:Graph1}. The NI controllers are the constant gains (-0.066, -0.02), \textit{X}$_{f}$=(0, 0) degree, \textit{Q}$_{i}$ = [1; -1], \textit{Q}$_{c}$= [0; -1], \textit{Q}$_{r}$= [1; 0], the \textit{X}$_{r}$ term is [90; 90] degree. These parameter settings are used to synchronize the yaw angle for the UGVs. The corresponding consensus result is depicted in Fig. \ref{fig:case2_yaw_angle}.

The simulation results have verified that our proposed algorithm can overcome the problem existing in time-varying formation control. Once the expected relative distances on the x and y axes are able to be altered at each vertex of the rectangular trajectory, our robots will take approximately 2 seconds to reach the new distance offset and to stabilize around this value. Additionally, the consensus of robots is performed successfully at each vertex of rectangle owing to our yaw angle consensus approach.

\section{The Geometry Method to Avoid Obstacles}

After verifying the feasibility of our NI varying formation control method in a simulation environment, it is necessary to indicate a new obstacle avoidance algorithm for the two avoiding cases: one obstacle alone and two obstacles in opposite sides.
 
\subsection{Identification of obstacles and robots}

Each robot is assumed to be equipped with a sensor (e.g. RGB-Depth camera) to determine the real obstacle boundaries. In a cluttered environment, the shape of real obstructions is intricate. Therefore, other studies, which compute the escape angle via actual border of the obstructions, often show the results within acceptable limits. Similarly, other approaches are implemented in a simple environment in which only some obstacles of simple shape exist. We solve this problem via two steps: (1) finding the center point of the real boundary parts within the Field of View (FOV) of the camera by using integral formula and geometric decomposition, and (2) generating a virtual circle that surrounds the recognized obstruction. The radius of this circle is equal to the distance between the center point and the furthest point of the observed obstacle boundary. By using the virtual circles to cover the obstacle boundaries, any computation based on the circle border line is more accurate. Each time the robot goes through the identified area, the calculating process will be repeated. 

\begin{figure}
	\centering
	\includegraphics[width=0.45\linewidth]{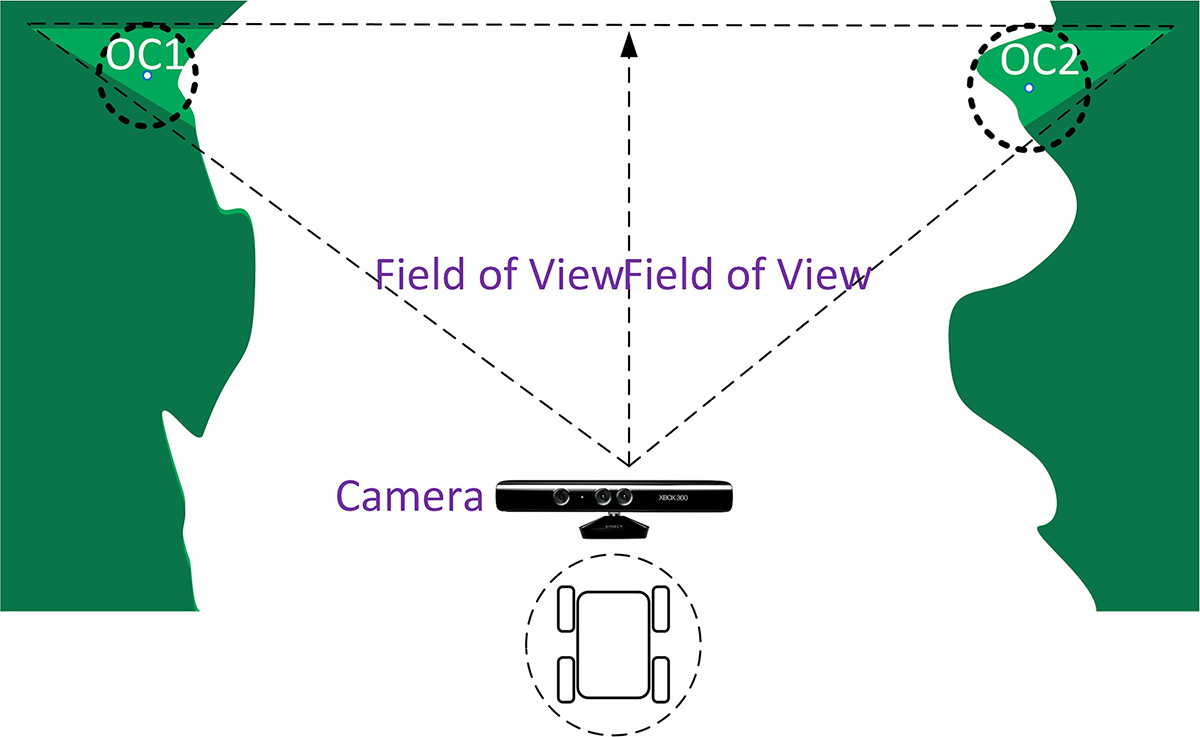}
	\caption{Generating the obstacle circles within the FOV of camera.}
	\label{fig:identification_obstacle}
\end{figure}

In Fig. \ref{fig:identification_obstacle}, the light-green colored regions which have polygon shapes are the overlapping ones between the camera's FOV and the top sections of the obstacle. Moreover, OC1 and OC2 are the geometric centers of the two polygons mentioned above and are also the centers of virtual circles which represent the viewable obstacle regions detected by the camera.

Occasionally, the obstacle boundary within the observation view is a discontinuity, which results in small spaces being created between the obstacles. It would be unacceptable if the avoiding algorithm drives our robot to pass through these narrow gaps. To overcome this obstruction, we group or separate these obstacles via the following conditions:

\begin{equation}
\begin{cases}
group$-$obstacle &\text{if dis12 $<$ d+r1+r2}, \\
separate$-$obstacle &\text{if dis12 $>=$ d+r1+r2}.
\end{cases}
\end{equation}

where \textit{dis12} is the distance between the two obstacle centers. \textit{r1} and \textit{r2} are the radii of the obstacle circle 1 and 2 respectively. \textit{d} is the diameter of the robot circumscribed circle.

\begin{figure}
	\centering
	\includegraphics[width=0.6\linewidth]{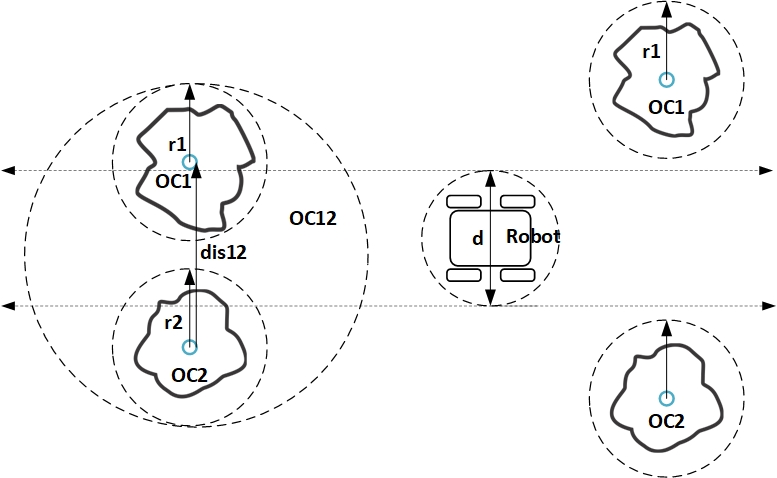}
	\caption{Necessary and sufficient conditions to group or separate obstacles.}
	\label{fig:group_obstacle}
\end{figure}

\begin{figure}
	\centering
	\includegraphics[width=0.55\linewidth]{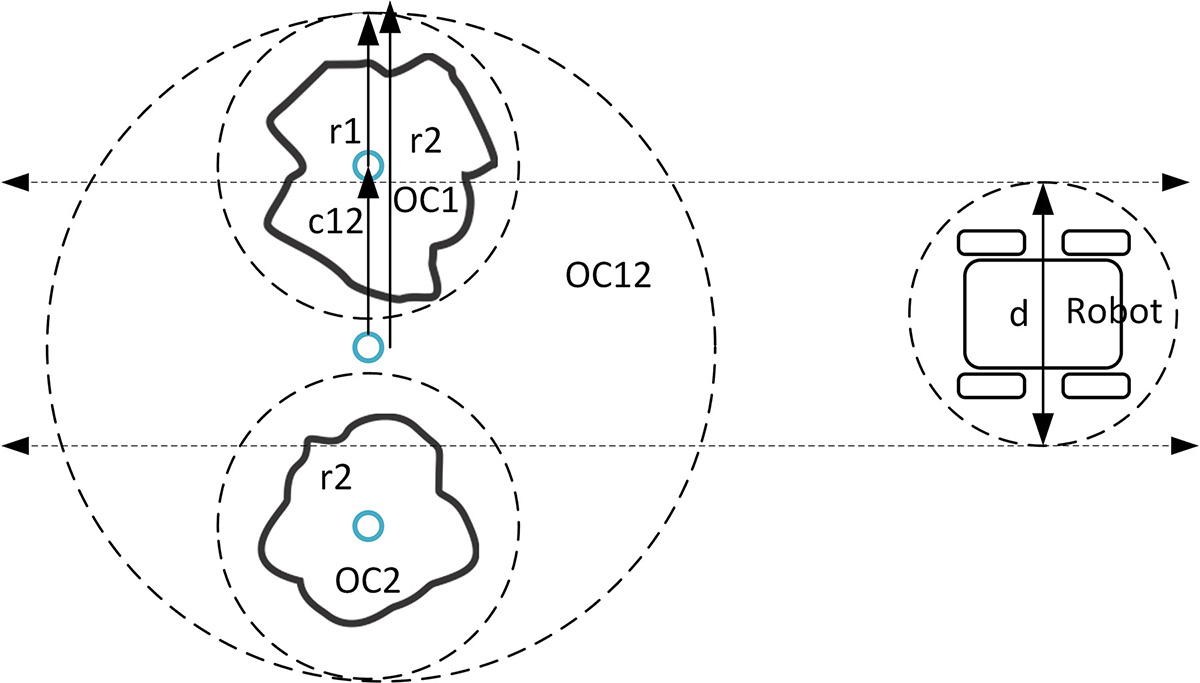}
	\caption{New obstacle circle design}
	\label{fig:new circle}
\end{figure}

As presented in Fig. \ref{fig:group_obstacle}, if the distance between the two centers of OC1 and OC2 is less than the diameter of the robot's virtual circle plus the two radii, a grouping is executed; otherwise, the separation is applied. The center coordinate and radius of the grouped circle OC12 is computed as shown in Fig. \ref{fig:new circle}. 

These computations can be extended for multi-obstacles situations if there are more than two obstacles whose locations satisfy the grouping condition. However, in case of using a laser rangefinder sensor; e.g., 3D LiDAR, which brings powerful 360-degree and 6-meter distance sensing capabilities, the obstacle circle radius and the number of grouped obstacles must be limited. The primary reason is that the larger area of the obstacle boundary this sensor can recognize, the smaller
the space between obstacles which the robots can bypass is.

Finally, by considering the size of each robot in our obstacle avoidance approach, each robot is also geometrically circumscribed by a virtual circle. Once the robot's virtual circle intersects with the obstacle circle, a potential collision is detected.
 
\subsection{Avoiding one obstacle}

We obtain the necessary and sufficient conditions of this mode as follows:

\begin{equation}
mode = \begin{cases}
1 &\text{if } ob \in FOV, ob \ni OB,  \\
    &\text{if } SP_{1} or SP_{2} \in SD_{1}SD_{2},\\
    &\text{if } dis(OC_{1}OC_{2}) > \frac{FOV}{2},\\
0 &\text{Otherwise}.
\end{cases}
\end{equation}

where \textit{SP1} and \textit{SP2} denote the coordinates of two projection points projected from the center line of two opposite obstacles. \textit{SD1} and \textit{SD2} illustrate the intersection points between the robot virtual circles and the center lines of two opposite robots. \textit{FOV} is the Field of View of a camera. \textit{OC1} and \textit{OC2} present the center location of each obstacle while \textit{SC1} and \textit{SC2} present the center location of each robot.

\begin{figure}
	\centering
	\includegraphics[width=0.69\linewidth]{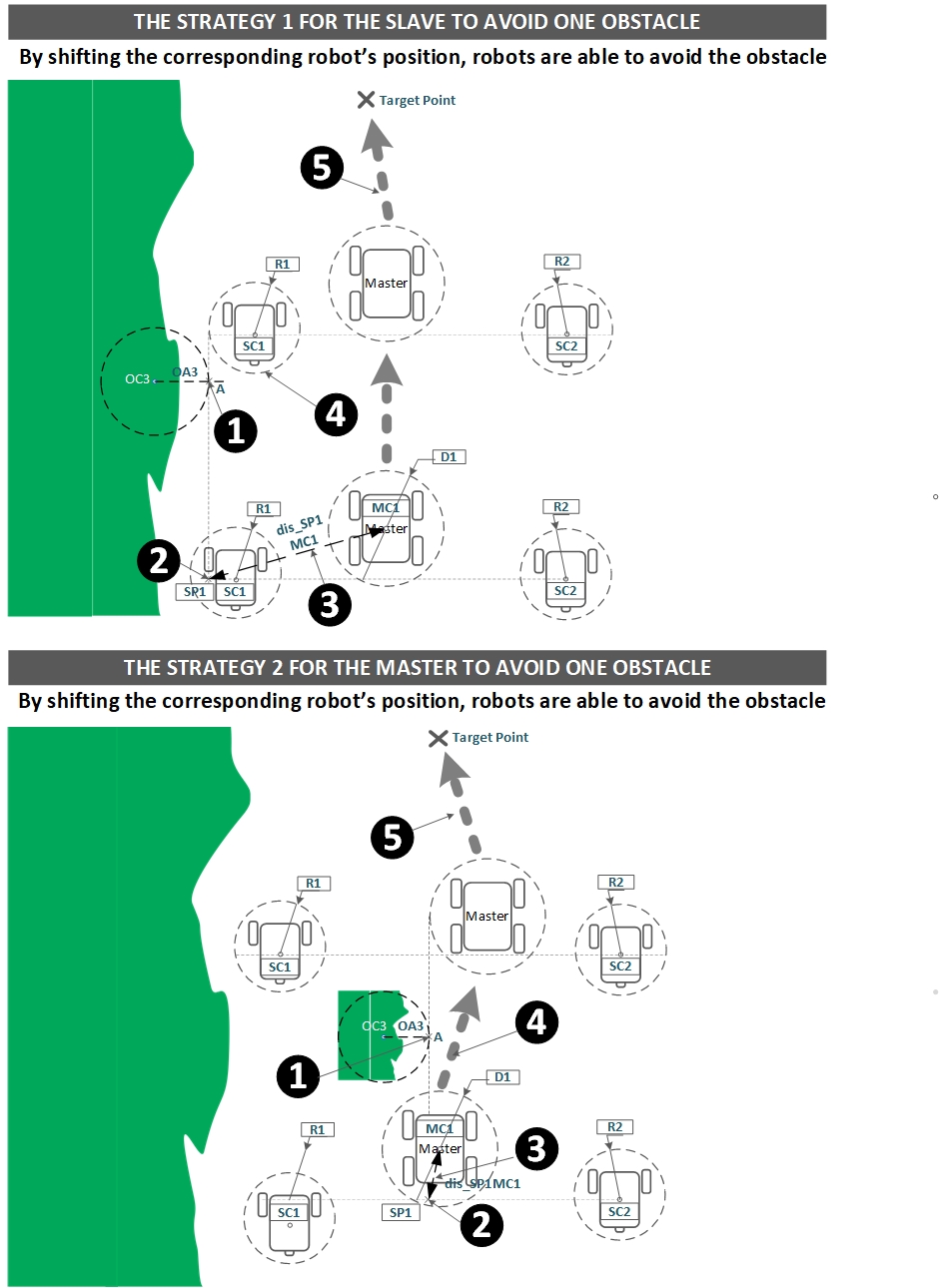}
	\caption{The two strategies for one obstacle-avoidance problem.}
	\label{fig:strategy11}
\end{figure}

After the requirements of the one-obstacle avoidance mode are determined, two strategies to transform the robots' formation will be designed. When one of the slaves predicts a potential collision with the obstacle (Strategy 1) via the distance between the obstacle and the robot, this robot will calculate the updated relative distance from the master position and will execute the avoiding process with the following steps: (1) calculating an intersection point \textit{A} between the obstacle circle and \textit{OA3}; (2) finding a projected point \textit{SP1} on \textit{SC1-SC2} line segment; (3) recomputing the relative distance \textit{dis$\_$SP1MC1} between the master and the corresponding slave; (4) the slave travels into the safe place; and (5) after avoiding process, the three robots reach the destination. Similarly, when the master detects a potential collision with the obstacle (Strategy 2), the methods for determining the projected point \textit{SP1} are the same as those mentioned in Strategy 1. However, in the third step (3), the \textit{SP1} coordinate is used as the position reference for the master while the slaves attempt to maintain their former orientation by storing its last position on the horizontal axis. All processes for two strategies are depicted as in Fig. \ref{fig:strategy11}.

\subsection{Avoiding two facing obstacles}

Regarding the evasion for obstacles located at two opposite sides, there are three situations which are mathematically illustrated as below:

\begin{equation}
mode = \begin{cases}
2 &\text{if } dis(SP_{1}SP_{2}) > dis(SC_{1}SC_{2}) + R{2} + R{3} (1),\\
&\& dis(OC_{1}OC_{2}) < \frac{FOV}{2} (1),\\
&\text{if } D{1} + R{1} + R{2}< dis(SP_{1}SP_{2}),\\
&\& dis(SP_{1}SP_{2}) < dis(SC_{1}SC_{2}) + R{2} + R{3} (2),\\
&\& dis(OC_{1}OC_{2}) < \frac{FOV}{2} (2),\\
&\text{if } D{1} < dis(SP_{1}SP_{2})(3),\\
&\& dis(SP_{1}SP_{2}) < D{1} + R{2} + R{3} (3),\\
&\& dis(OC_{1}OC_{2}) < \frac{FOV}{2} (3),\\
0 &\text{Otherwise}.
\end{cases}
\end{equation}

where \textit{D{1}} is the master robot's virtual circle diameter

In Case 1 (1), the three robots are able to maintain the configured formation and move to the target without being blocked by any obstruction due to a large clearance between two obstacles. In Case 2 (2) and Case 3 (3), the three robots are forced to alter their formation pattern while there are two opposite obstacles blocking their path and the relative distance between the master and the obstacles is 1 meter. The former will modify the distance between robots while the latter will shift from the original formation to the queuing formation. In the scope of this paper, we will concentrate on solving the problems arising from the first and second cases.

\begin{figure}
	\centering
	\includegraphics[width=0.7\linewidth]{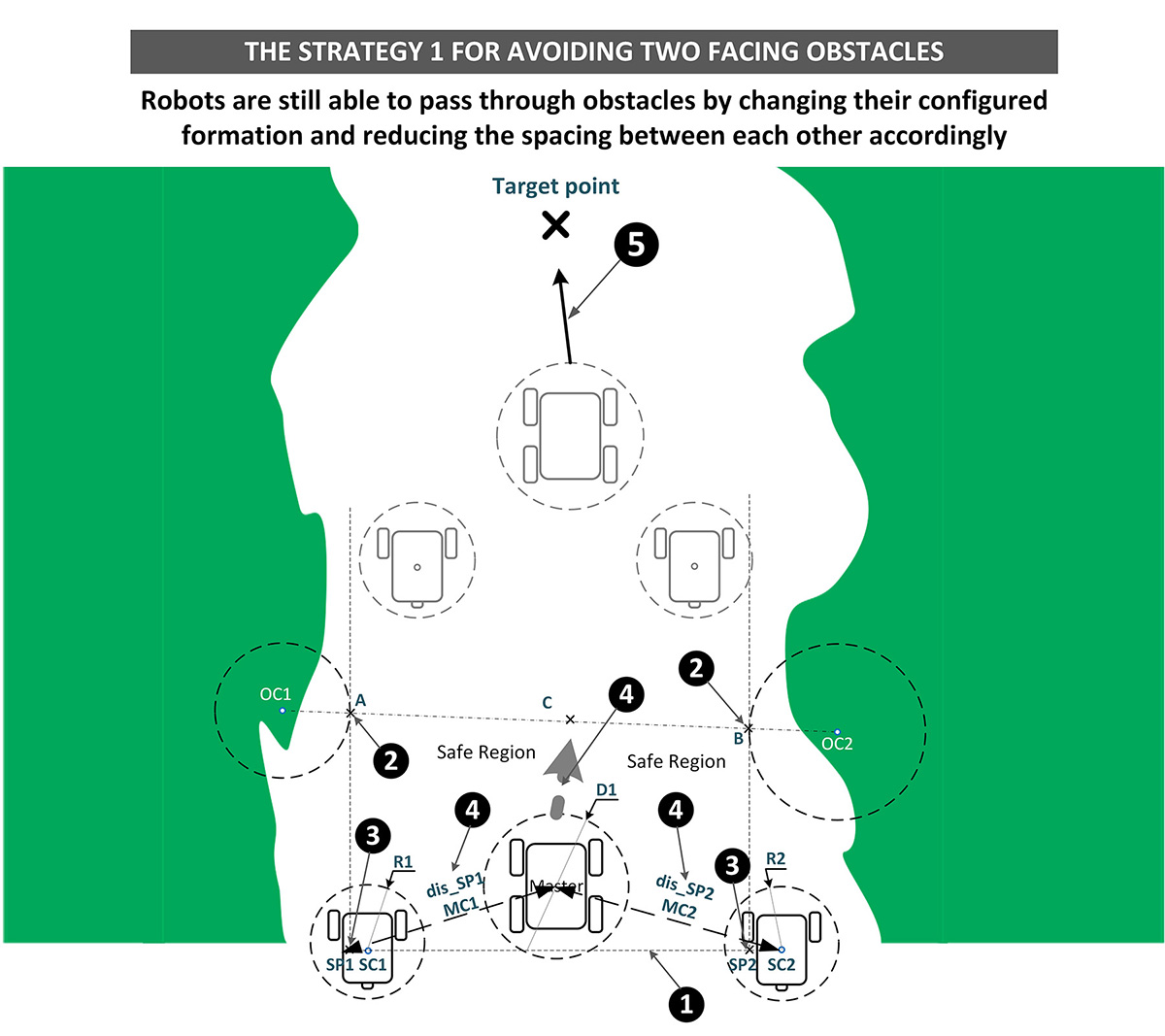}
	\caption{The strategy for two-facing-obstacles avoidance problem.}
	\label{fig:strategy2}
\end{figure}

When the conditions of Case 1 are met, no action is required by the robots. However, when the conditions of Case 2 are satisfied, our robots will be guided following five steps as shown in Fig. \ref{fig:strategy2}: (1) creating the centers line \textit{SC1-SC2} between the two slaves; (2) finding the two intersection points \textit{A} and \textit{B} between the obstacle's boundary circles and the connecting line \textit{OC1-OC2}; (3) determining their two projected points \textit{SP1} and \textit{SP2} on the \textit{SC1-SC2} line; (4) computing the new desired relative distance that the two slaves must follow while the master robot is driven to the middle point \textit{C} of \textit{OC1-OC2}; and (5) the former formation pattern will be restored right after the master robot moves out of FOV of the master's camera.

\section{Experimental Results}

Three experiments have been presented to illustrate the main results of
this research. Test 1 introduces the ability of invariant formation control via the NI consensus strategy applied to multi-vehicle systems. In contrast, Test 2 demonstrates the transformation of robots' time-variant formation pattern with the main purpose of keeping away from obstacles in two various situations: one obstacle and two group of obstacles in parallel.  The final test is conducted for an overall situation in which obstacles are placed randomly.

\begin{figure}
	\centering
	\includegraphics[width=0.25\linewidth]{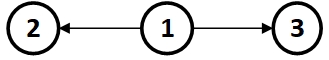}
	\caption{Communication Graph for Test 2.}
	\label{fig:Graph2}
\end{figure}

\begin{figure}
	\centering
	\includegraphics[width=0.95\linewidth]{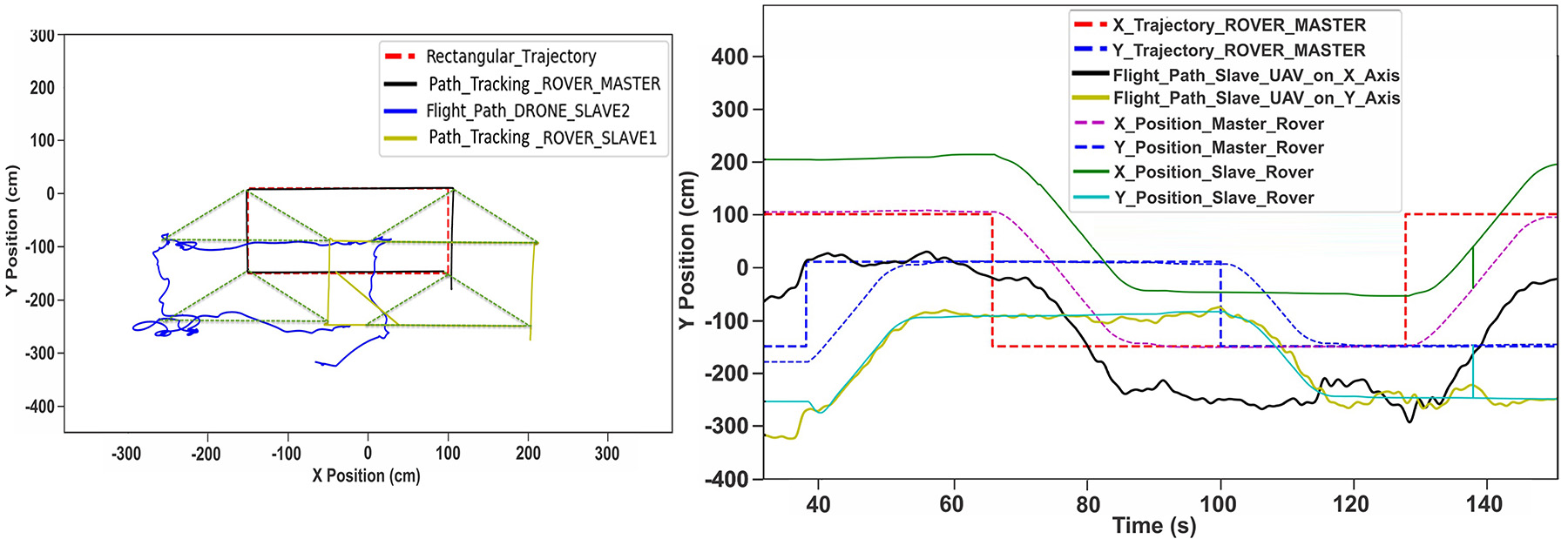}
	\caption{Experimental results for the triangular formation control based on a position consensus involving the two UGVs and the UAV.}
	\label{fig:test1}
\end{figure}

\begin{figure}
	\centering
	\includegraphics[width=0.65\linewidth]{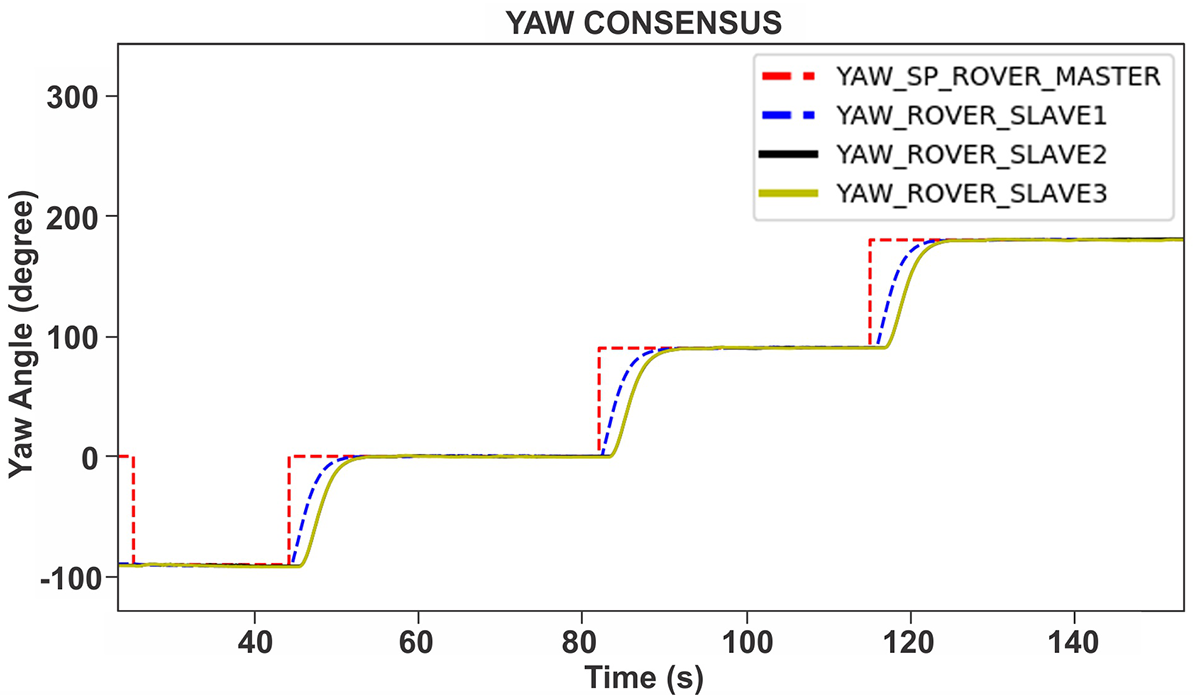}
	\caption{Experimental results for a yaw angle consensus involving two UGVs.}
	\label{fig:Figure_11}
\end{figure}

\textbf{Test 1}: Test 1 aims at validating the efficiency of both consensus and formation control for multi-robot systems, including two UGVs and one UAV. The master UGV (system 1 in Fig. \ref{fig:Graph2}) will receive the waypoints on the rectangular trajectory path while the slaves, involving one UGV (system3) and one UAV (system2), follow the master and maintain a relative distance, determined by the pre-defined triangular formation. The geometrical distance between the master and each slave is 100 cm. This term is configured by the \textit{X}$_{f}$ parameter in the enhanced NI formation control diagram. The communication topology is given in Fig. \ref{fig:Graph2}. The NI controllers are the constant gains (-0.0028, -0.0028, -0.7, -0.7, -0.0069, -0.0069), \textit{X}$_{f}$=(100, -100, 100, 100, 0, 0) cm. The \textit{X}$_{r}$ parameter is the location matrix for the four vertexes of the rectangle. The coordinates of four vertexes are [100, -150], [100, 10], [-150, 10], [-150, -150]. Moreover, the yaw angles of the two UGVs are also synchronized with each other using the same approach. The NI controllers used for yaw angle consensus are the constant gains (-0.0056, -0.0107). Similarly, \textit{X}$_{f}$=(0, 0) degrees. The \textit{X}$_{r}$ parameter is referred to as the desired angle setpoints set and is calculated using the rotation matrix and the next vertex location. All experimental results are shown in Fig. \ref{fig:test1} and Fig. \ref{fig:Figure_11}.

\begin{figure}
	\centering
	\includegraphics[width=0.75\linewidth]{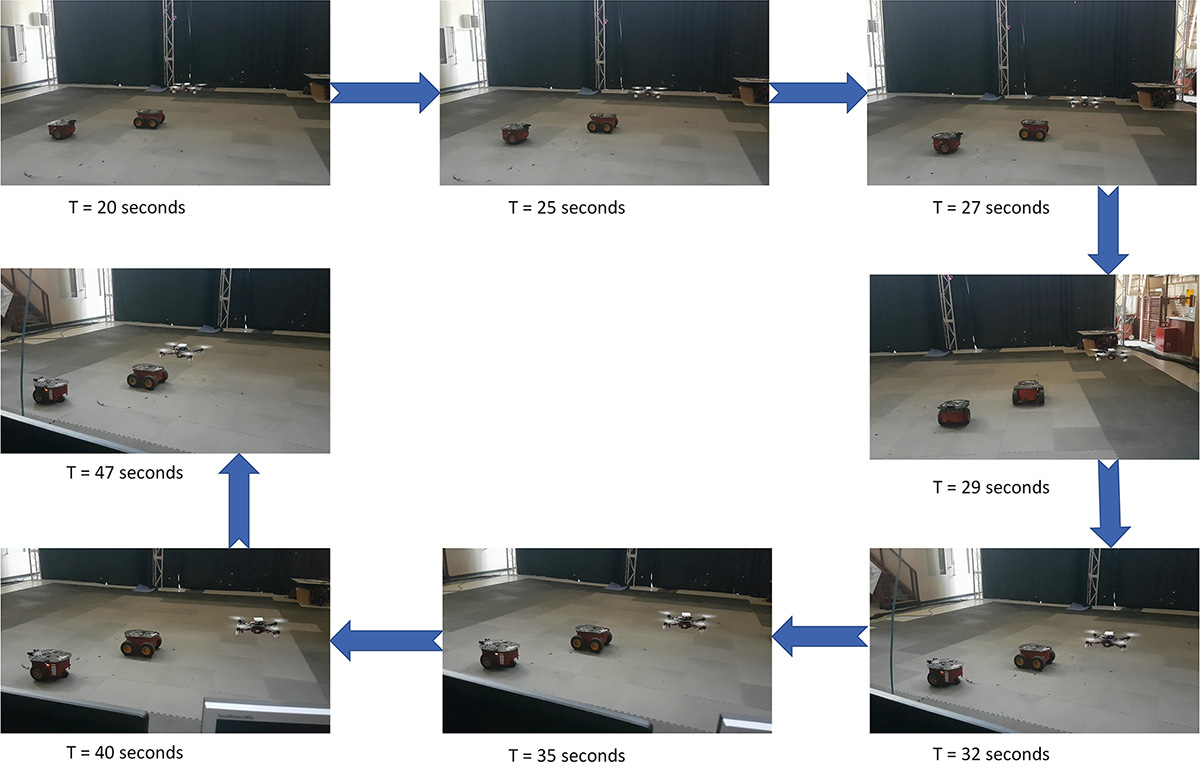}
	\caption{Formation control experimentation. The eight images represent the linear displacement of the triangle pattern generated by the two UGVs and the UAV. The robots preserve this pattern and travel along the rectangular path simultaneously.}
	\label{fig:lab.jpg}
\end{figure}

For more details, Fig. \ref{fig:lab.jpg} illustrates graphically a triangle shape-shifting in space produced by virtual connections made by the central poses of the two UGVs and the UAV (the different line types and colors) based on the enhanced NI formation control strategy with varying time steps. 

As seen in Fig. \ref{fig:test1} and Fig. \ref{fig:lab.jpg}, good formation and consensus for a group of UAV and UGVs are guaranteed. Based on the desired relative distance information with the master position, the two slaves, including the UAV and the UGV, together with the master UGV, have generated an inverted-V-shaped formation (green-colored lines) and moved along the rectangular paths (green, blue and black-colored lines). The maximum error of position is approximately 15 cm for the UAV and 7 cm for the UGVs. 

\begin{figure}
	\centering
	\includegraphics[width=0.95\linewidth]{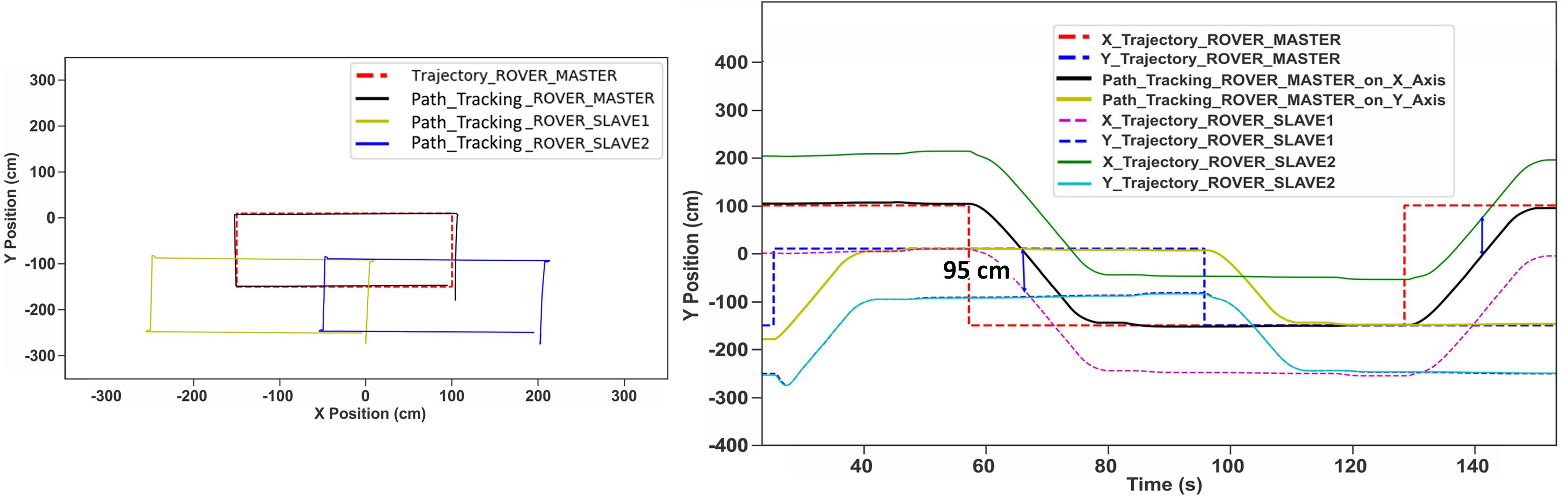}
	\caption{Experimental results for the triangular formation control based on a position consensus involving the three UGVs.}
	\label{fig:test2}
\end{figure}

\begin{figure}
	\centering
	\includegraphics[width=0.75\linewidth]{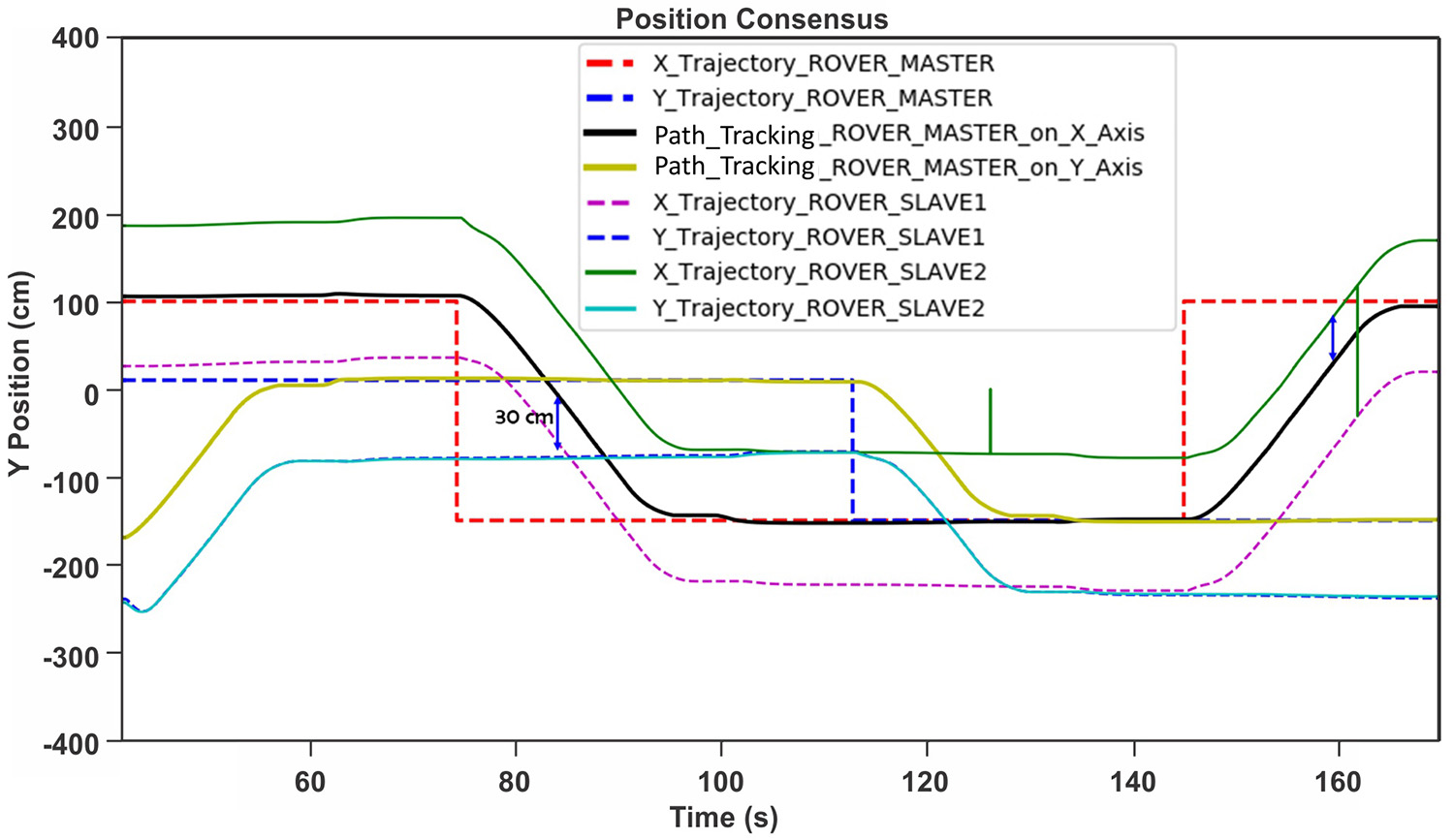}
	\caption{Position response of the three UGVs to the original NI consensus strategy.}
	\label{fig:Figure_72}
\end{figure}

Moreover, response time of systems under the control of the enhanced NI formation architecture is improved dramatically. In order to provide more clear proof of this improvement, the formation experiments using our former NI architecture for three UGVs \cite{Tran17} are conducted in a similar way as those in Test 1. Fig. \ref{fig:test2} and Fig. \ref{fig:Figure_72} show the performances of the three UGVs over time. Compared with the old architecture in stabilizing the desired distance of 100 cm between the UGVs, the actual relative distance between the UGVs in Fig. \ref{fig:test2} is up to 95 cm while that in Fig. \ref{fig:Figure_72} is only 30 cm. This verifies that the enhanced NI architecture for formation control is better than the ordinary one. 

Finally, the rectangular trajectory of the two UGVs in Fig. \ref{fig:test1} is extremely smooth, especially at the four vertexes. Such performance is achieved thanks to strong yaw angle consensus as shown in Fig. \ref{fig:case2_yaw_angle} and Fig. \ref{fig:Figure_11}.

\textbf{Test 2}: Two avoidance schemes mentioned in Section VIII are implemented separately in an indoor environment using the UAV-UGV/UGVs systems. These systems are integrated in our NI time-varying formation control protocol. In experiment 1, the master UAV attempts to evade the obstacle by changing its waypoints. Meanwhile, the two slave UGVs maintain their pre-defined formation. Different from the first test, experiment 2 concentrates on how the three UGVs can overcome the two groups of obstacles located at two opposite sides.

All settings of two experiments are as per those of Test 1. The radius of the robot circumscribed circles is 32 cm, and that of the obstacle's boundary circles is 35 cm. The FOV of the camera is roughly 2.2 m. The coordinate of the destination point in the experiment 1 is [-100, 170] while it is  [-100, 220] in the experiment 2.

\textbf{Test 3}: In the final experiment, an uncertain environment, containing two opposite-side obstacles and a single obstacle, is explored by a team of 3 UGVs. The target is assigned at the coordinate [-100, 240]. The task of this team is to preserve the triangle-formation as best as possible while avoiding unexpected obstacles and moving to the target simultaneously. 

\begin{figure}
	\centering
	\includegraphics[width=0.8\linewidth]{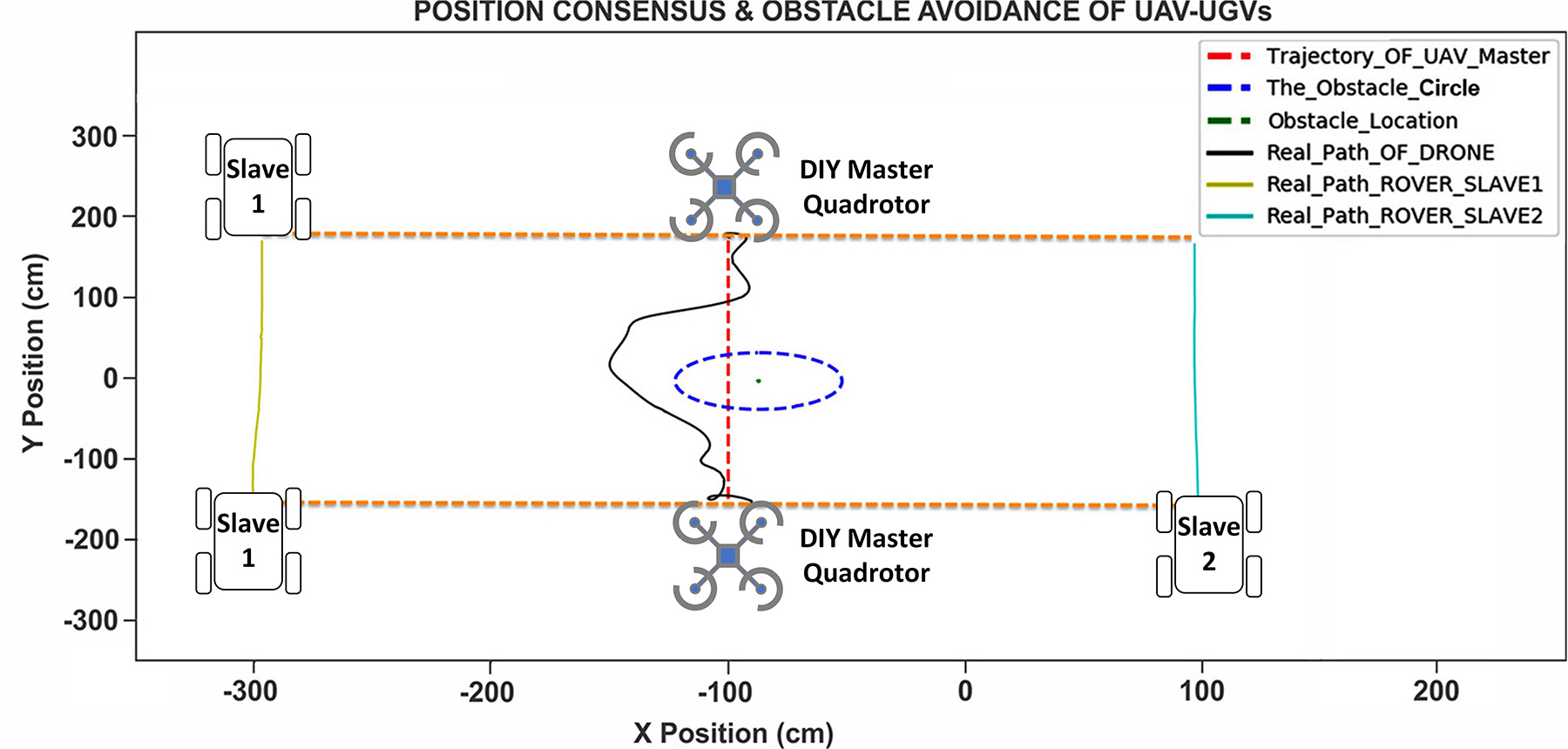}
	\caption{Experimental results for the varying horizontal-line formation and obstacle avoidance control based on the NI consensus architecture involving the leader UAV and the two slaves UGV.}
	\label{fig:strategy2_result21}
\end{figure}

\begin{figure}
	\centering
	\includegraphics[width=0.8\linewidth]{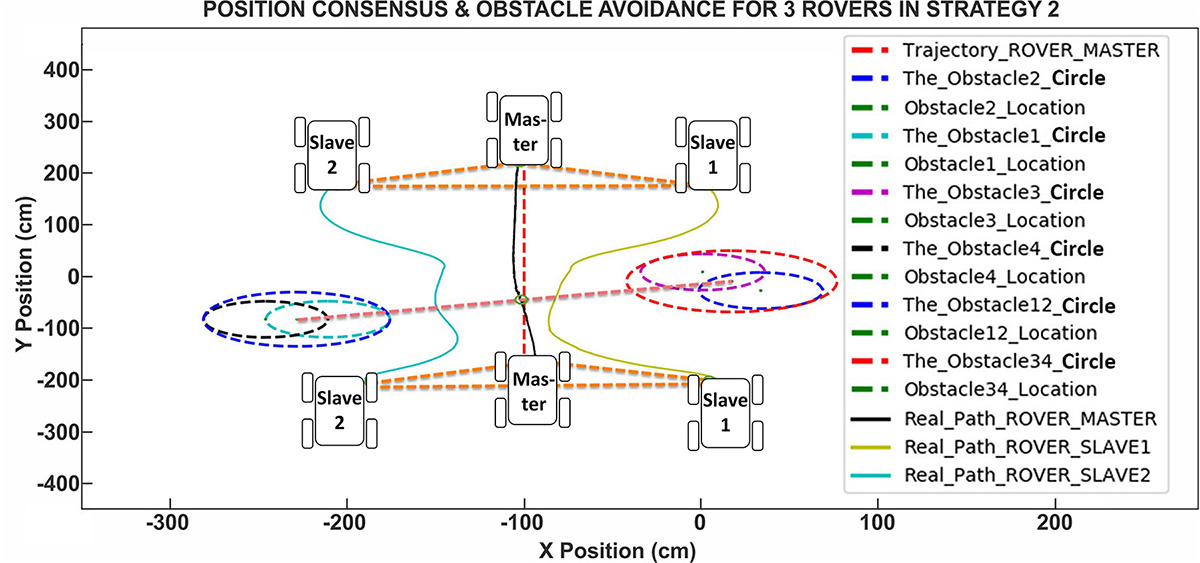}
	\caption{Experimental results for the triangle varying formation and obstacle avoidance control based on the NI consensus architecture involving the three UGVs.}
	\label{fig:strategy22}
\end{figure}

\begin{figure}
	\centering
	\includegraphics[width=0.8\linewidth]{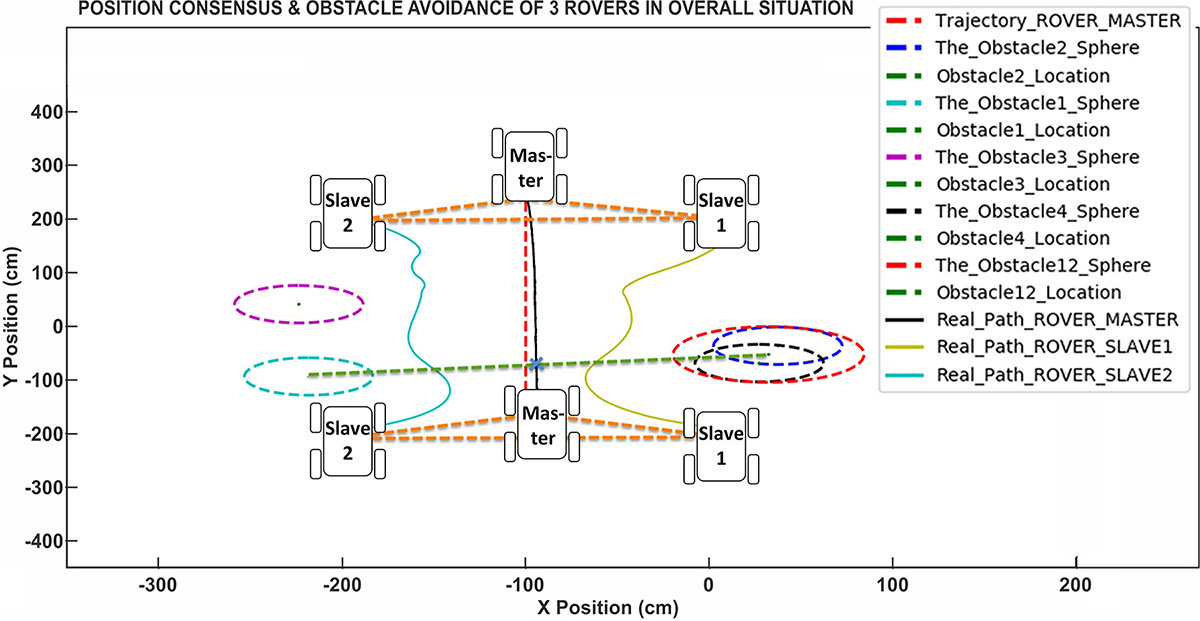}
	\caption{Experimental results for the triangle varying formation and obstacle avoidance control based on the NI consensus architecture involving the three UGVs in an uncertain environment.}
	\label{fig:overall_situation1}
\end{figure}

Fig. \ref{fig:strategy2_result21} shows the significant results of the straight-line shaped formation and the master-obstacle collision avoidance. The initial objectives of this experiment are achieved. The straight-line shaped formation is preserved perfectly by the two UGVs even though the leader must avoid the obstacle. Similarly, Fig. \ref{fig:strategy22} indicates that the three UGVs are driven appropriately according to the first strategy of avoiding two groups of opposite obstacles (Section VIII) while maintaining their triangular formation. 

Following the success of Test 2, Test 3 is conducted without any constraints. The three UGVs easily pass through any obstacles on their path and reach the target despite the two facing arrays of multiple obstacles and a single obstacle are placed randomly in indoor environment. 

In each test, the position consensus errors made by the robots are approximately 10 cm on the x and y axes. Since there are not any oscillations occurring in the real robot trajectories, it is pointed out that our formation methodology is better than that of authors in \cite{Dai15}. Besides, our approach also pays attention to the size of each robot. It is noted that the distance between the real avoiding path of all examined robots and the obstacle circle \textit{OC12} is roughly 31 cm with respect to the desired radius of 32 cm as shown in Fig. \ref{fig:strategy2_result21} to Fig. \ref{fig:overall_situation1}. Moreover, the formation regrouping takes place smoothly via the robots' steering behavior. The boundary of the restoration trajectory is the total of the robot radius and the border line of the obstacle's virtual circle. Finally, thanks to the relative distance calculation via the boundary of the obstacle's virtual circle, the safe path generated is better and smoother than that of the traditional GOACM method which entirely depends on the real boundary of the complex-shaped obstacle.

As seen in Fig. \ref{fig:test1} and Fig. \ref{fig:strategy2_result21}, there is no problem when the master is either the UAV or the UGV although two systems have very different dynamics. The UAV's outer-loop velocity controller sends attitude commands to the inner-loop attitude controller which commands the motors driving the rotors. In contrast, the UGV's velocity controller controls the velocity of four wheels directly. This control is achieved by converting the formation control into the velocity stabilization of each robot on the x and y axes via the NI-systems control algorithm.  

\section{Conclusion}

A time-varying formation control scheme based on the enhanced NI systems consensus theory for two popular types of mobile robots (UAVs-UGVs) has been introduced in this paper. In this consensus problem, the master corrects its state in space upon receiving waypoints from the path planner, while the slaves predict the next position of the master and will respond to its movement with an offset distance that is determined by the formation planner. In the obstacle avoidance problem, comprehensive situations are considered, and the experiments in simulation and real tests present the positive results of the proposed method compared to the recent studies.

Some position errors at the sharp corners are experienced by the UAV as shown in Fig. \ref{fig:test1}. The distance reference with the leader UGV when this leader rotates at a sharp corner and the lack of a real velocity sensor are the major causes for these errors. In future, to handle these problems, the measured distance between the master UGV and the slave UAV should be neglected during the rotational motion of the UGVs. To deal with the second issue, the UAV platform should be equipped with a velocity sensor (e.g. optical flow based sensor) to measure the real velocities on the x and y axes. 

In future work, other applications generated by formation control architecture such as distributed consensus protocols, leaderless formation or group control of UAVs-UGVs using NI theory will also be studied; notably, the robot alignment, the V-shape formation control problem or the actual formation of cooperative groups.

% trigger a \newpage just before the given reference
% number - used to balance the columns on the last page
% adjust value as needed - may need to be readjusted if
% the document is modified later
%\IEEEtriggeratref{8}
% The "triggered" command can be changed if desired:
%\IEEEtriggercmd{\enlargethispage{-5in}}

% references section

% can use a bibliography generated by BibTeX as a .bbl file
% BibTeX documentation can be easily obtained at:
% http://mirror.ctan.org/biblio/bibtex/contrib/doc/
% The IEEEtran BibTeX style support page is at:
% http://www.michaelshell.org/tex/ieeetran/bibtex/
%\bibliographystyle{IEEEtran}
% argument is your BibTeX string definitions and bibliography database(s)
%\bibliography{IEEEabrv,../bib/paper}
%
% <OR> manually copy in the resultant .bbl file
% set second argument of \begin to the number of references
% (used to reserve space for the reference number labels box)
% Can use something like this to put references on a page
% by themselves when using endfloat and the captionsoff option.

% You can push biographies down or up by placing
% a \vfill before or after them. The appropriate
% use of \vfill depends on what kind of text is
% on the last page and whether or not the columns
% are being equalized.

%\vfill

% Can be used to pull up biographies so that the bottom of the last one
% is flush with the other column.
%\enlargethispage{-5in}

% that's all folks
\end{document}